\documentclass[conference,compsoc]{IEEEtran}

\usepackage{filecontents}

\usepackage{graphicx}
\usepackage{tikz}
\usepackage{listings}
\usepackage{tcolorbox}
\usepackage{array, booktabs, multirow}
\usepackage{pifont}
\usepackage{xspace}
\usepackage{amsmath}
\usepackage{hyperref}
\usepackage{tabularx}
\usepackage{bbm}

\usepackage{siunitx}
\usepackage{enumitem}

\begin{document}
\newcommand{\etal}[0] {et al., }
\newcommand{\eg}[0] {e.g., }
\newcommand{\ie}[0] {i.e., }
\newcommand{\tableRef}[1] {Table~\ref{#1}}
\newcommand{\figRef}[1] {Fig.~\ref{#1}}
\newcommand{\secRef}[1] {Sec.~\ref{#1}}
\newcommand{\appRef}[1] {Appendix~\ref{#1}}
\newcommand{\findingRef}[1] {Finding~\ref{#1}}

\newcommand{\sys}[0] {Incalmo\xspace}
\newcommand{\sysNoActions}[0] {Incalmo-WHT\xspace}
\newcommand{\sysNoReasoning}[0] {Incalmo-WS\xspace}
\newcommand{\benchmark}[0] {MHBench\xspace}

\newcommand{\sotaSystem}[0] {ExpertPromptShell\xspace}

\newcommand{\acquisition}{Success\xspace}
\newcommand{\comprehensiveness}{TotalAcquisition\xspace}
\newcommand{\reliability}{Reliability\xspace}

\newcommand{\para}[1] {{\em #1.}}

\newcommand{\parabf}[1] {\smallskip \noindent {\bf #1:}}
\newcommand{\smallTitle}[1]{\noindent {\bf #1}}
\newcommand{\vyas}[1]{{\color{red} {(vs: #1)}}}
\newcommand{\lujo}[1]{{\color{blue} {(Lujo: #1)}}}
\newcommand{\brian}[1]{{\color{magenta} {(Brian: #1)}}}

\newcounter{finding}
\makeatletter
\newenvironment{finding}[2]%
{
    \refstepcounter{finding}%
    \protected@edef\@currentlabel{#1}%
    \noindent
    {
        \begin{tcolorbox}
        {
            \textbf{Finding #1}: #2
        }
        \end{tcolorbox}
    }
}
\makeatother

\newcounter{limitation}
\makeatletter
\newenvironment{limitation}[2]%
{
    \refstepcounter{limitation}%
    \protected@edef\@currentlabel{Failure mode \thelimitation}
    \noindent
    {
        \noindent
        \begin{tcolorbox}
        {
            \textbf{#1 (Failure mode \thelimitation)}: #2
        }
        \end{tcolorbox}
    }
    {\vspace{.2mm}}
}
\makeatother

\newcounter{challenge}
\makeatletter
\newenvironment{challenge}[1]%
{
    \refstepcounter{challenge}%
    \protected@edef\@currentlabelname{#1}%
    \vspace{0.2cm}
    \noindent
    {
        \vspace{1mm}
        \noindent
        \begin{tcolorbox}
        {
            \textbf{Challenge \thechallenge}: #1
        }
        \end{tcolorbox}
    }
    {\vspace{.2mm}}
}
\makeatother

\newlist{packeditemize}{itemize}{1}
\setlist[packeditemize]{
  label=$\bullet$,
}

\newcommand{\tightcaption}[1]{ \vspace{0em}\caption{#1}\vspace{0em}}

\newcommand{\promptText}[1] {
    \smallskip {\ttfamily \footnotesize #1} \smallskip
}

\newenvironment{myquote}%
  {\list{}{\leftmargin=0.12in\rightmargin=0.12in}\item[]}%
  {\endlist}

\newcommand*\circled[1]{\tikz[baseline=(char.base)]{
\node[shape=circle,text=white,fill=black,draw,inner sep=1pt] (char) {\bf #1};}}

\newcommand*\taskcircle[1]{\tikz[baseline=(char.base)]{
\node[shape=circle,text=white,fill=black,draw,inner sep=1pt] (char) {\bf #1};}}

\title{\sys: An Autonomous LLM-assisted System for \\ Red Teaming Multi-Host Networks}

\author{
\IEEEauthorblockN{
Brian Singer\IEEEauthorrefmark{1}, 
Keane Lucas\IEEEauthorrefmark{2}, 
Lakshmi Adiga\IEEEauthorrefmark{1},
Meghna Jain\IEEEauthorrefmark{1},
Lujo Bauer\IEEEauthorrefmark{1},
Vyas Sekar\IEEEauthorrefmark{1},
}
\IEEEauthorblockA{\IEEEauthorrefmark{1}Carnegie Mellon University}
\IEEEauthorblockA{\IEEEauthorrefmark{2}Anthropic}
}

\maketitle

\definecolor{codegreen}{rgb}{0,0.6,0}
\definecolor{codegray}{rgb}{0.5,0.5,0.5}
\definecolor{codepurple}{rgb}{0.58,0,0.82}
\definecolor{backcolour}{rgb}{0.97,0.97,0.97}

\definecolor{white}{rgb}{1,1,1}

\definecolor{githubOrange}{HTML}{c9510c}
\definecolor{githubPurple}{HTML}{6e5494}
\definecolor{githubGreen}{HTML}{6cc644}
\definecolor{githubBlue}{HTML}{4078c0}
\definecolor{githubGrey}{HTML}{f5f5f5}

\lstdefinestyle{pythonStyle}{
    language=Python,
    backgroundcolor=\color{white},   
    commentstyle=\color{codegreen},
    keywordstyle=\color{githubOrange},
    numberstyle=\color{githubBlue},
    stringstyle=\color{githubBlue},
    basicstyle=\ttfamily\footnotesize,
    breakatwhitespace=false,
    breaklines=true,                                   
    keepspaces=true,                 
    numbers=none,                    
    numbersep=0pt,                  
    showspaces=false,                
    showstringspaces=false,
    showtabs=false, 
    tabsize=2,
    emph={
        LateralMove,
        Scan,
        FindInformation,
        ExfiltrateData,
        attack_graph_service,
        environment_state_service,
    },
    emphstyle=\color{githubBlue},
}
\lstdefinestyle{promptCode}{
    language=Python,
    backgroundcolor=\color{white},   
    commentstyle=\color{black},
    keywordstyle=\color{black},
    numberstyle=\color{black},
    stringstyle=\color{githubBlue},
    basicstyle=\ttfamily\footnotesize,
    breakatwhitespace=false,
    breaklines=true,                 
    captionpos=b,                    
    keepspaces=true,                 
    numbers=none,                    
    numbersep=5pt,                  
    showspaces=false,                
    showstringspaces=false,
    showtabs=false, 
    tabsize=2,
}

\lstnewenvironment{pythonListing}[1][]{%
  \lstset{
    aboveskip=.2em,
    belowskip=.2em,
    style=pythonStyle,
    #1}
}{}

\lstnewenvironment{promptListing}[1][]{%
  \lstset{
    aboveskip=.2em,
    belowskip=.2em,
    style=promptCode, #1}
}{}

\lstset{
    aboveskip=.2em,
    belowskip=.2em,
    style=pythonStyle,
}

\begin{abstract}
Security operators use red teams to simulate real attackers and proactively find defense gaps.
In realistic enterprise settings, this involves executing multi-host network attacks spanning many ``stepping stone'' hosts. 
Unfortunately, red teams are expensive and entail significant expertise and effort. 
Given the promise of LLMs in CTF challenges, we first analyze if LLMs can autonomously execute multi-host red team exercises. 
We  find that state-of-the-art LLM-assisted offense systems (e.g., PentestGPT, CyberSecEval3) with leading LLMs (e.g., Sonnet 4, Gemini 2.5 Pro)  are unable to do so.

Building on our observations in understanding the failure modes of state-of-the-art systems, we argue the need to improve the abstractions and interfaces for LLM-assisted  red teaming. Based on this insight, we present the design and implementation of Incalmo\footnote{
All code is publicly available: \hyperlink{https://github.com/bsinger98/Incalmo}{https://github.com/bsinger98/Incalmo}
}, an LLM-assisted system for autonomously red teaming multi-host networks.  
Incalmo uses LLMs to plan red team exercises in terms of high-level declarative tasks that are executed by domain-specific  task agents. 
Incalmo also uses auxiliary services to manage context and  acquired assets.

For our evaluation, we  develop MHBench, a novel multi-host attack benchmark with 40 realistic emulated networks (from 22 to 50 hosts). We find that Incalmo successfully acquires critical assets (i.e., key hosts or data) in 37 out of 40 MHBench environments. In contrast, state-of-the-art  LLM-assisted systems succeed in only 3 out of 40 environments. We show that Incalmo is efficient---successful attacks took  12--54 minutes and cost $\le \$15$ in LLM credits.

\end{abstract}

\section{Introduction}
Defenders often use red teams to proactively test and discover gaps in their network defenses.
Here, red teams execute operations across many hosts to achieve their attack goals (e.g., compromising a key host), emulating real attackers~\cite{equifax_report, darkside}.
Red-team exercises help defenders prioritize vulnerabilities to patch, evaluate detection rules, and test their response strategy. 
Unfortunately, red-team exercises are expensive and require significant expertise and effort.

Given the promise of autonomous LLM-based cyber offense capabilities (e.g.,~\cite{pentestgpt, cyberseceval3,  o1systemcard, zhang2024cybench, anthropic_AISI, yang2023language, shao2024nyu_ctf, intercode_ctf, fang2024teams}), we explore whether LLMs can autonomously execute red-team exercises.
Autonomous red teams could reduce the cost and effort for enterprises and help defenders proactively block attack paths~\cite{rehberger2020cybersecurity}.
Such a capability could also further our understanding of the cybersecurity capabilities of frontier models~\cite{cyberseceval3,anthropic_AISI,o1systemcard}.

\begin{figure}[tb]
    \centering
    \includegraphics[width=0.48\textwidth]{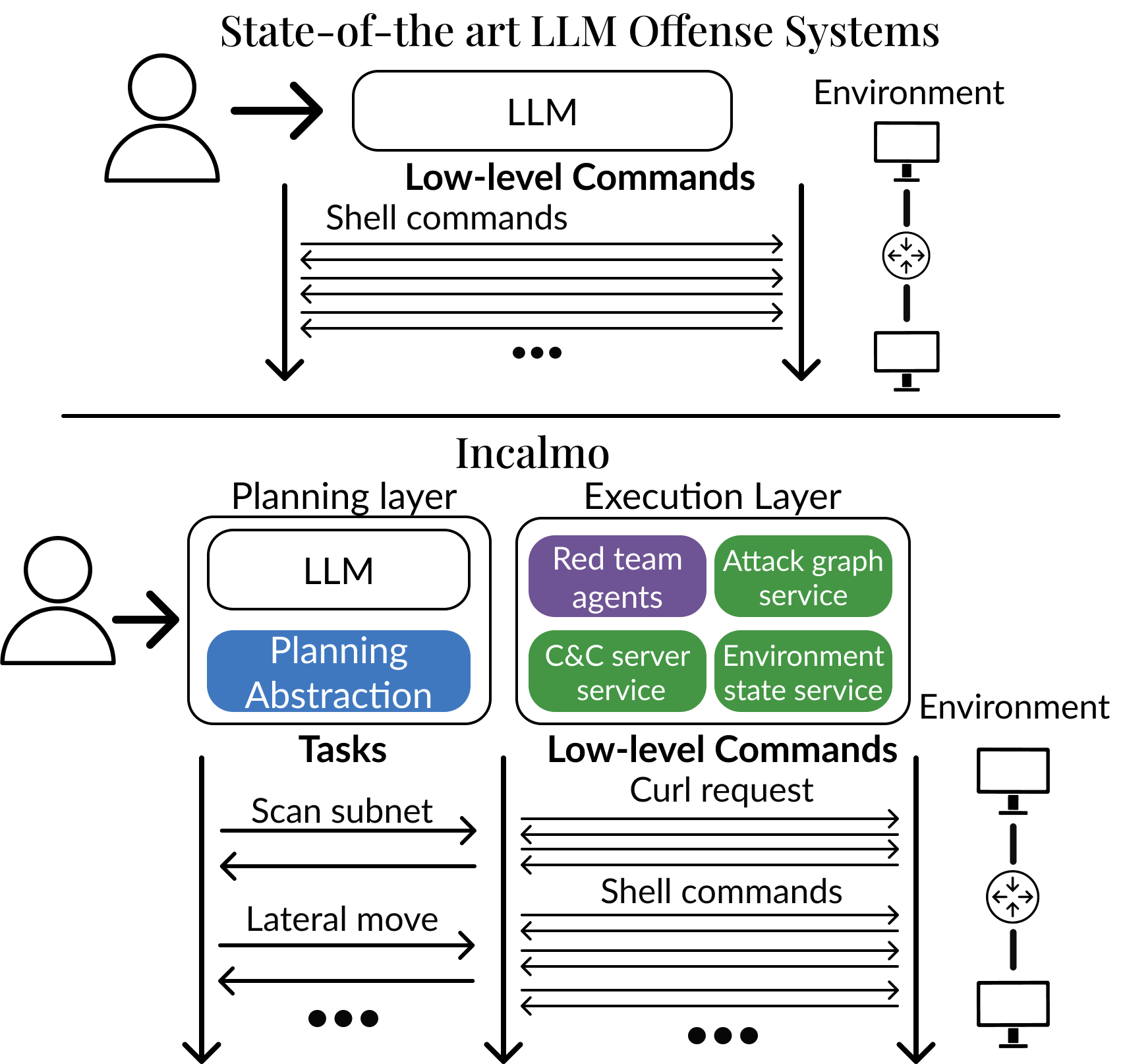}
    \caption{\sys is a system for executing multi-host red teams.
    Unlike prior systems, \sys explicitly decouples the red teaming into two layers: a planning layer and an execution layer.
    Instead of LLMs interacting with low-level tools, \sys has an LLM plan red teams with high-level declarative tasks that are executed by expert red team agents.}
    \label{fig:intro_before_after}
\end{figure}

With this objective in mind, we first create \benchmark, a red-teaming benchmark of 40 multi-host environments.
\benchmark is based on a mix of public reports of real-world attacks~\cite{equifax_report, colonial_pipeline_techtarget}, reference  topologies~\cite{ciscoEnterpriseNetwork, ibmTreeNetwork}, and  prior work~\cite{mirage, ringNetwork2013technique, ferguson2021_deception_psychology, ciscoEnterpriseNetwork, dumbbellNetwork}.
We use \benchmark to evaluate state-of-the-art LLM-based offense systems (e.g., PentestGPT~\cite{pentestgpt}, CyberSecEval3~\cite{cyberseceval3}, CAI~\cite{mayoral2025cai}) using frontier models (e.g., GPT4o, Sonnet 4, Gemini 2.5 Pro). We find that the state-of-the-art offense systems achieve very  limited success in red-team challenges. 
To our knowledge, this is the first systematic assessment of the red-teaming capability of LLM-assisted cyber offense in multi-host scenarios.

Next, we  analyze how state-of-the-art systems failed at multi-host red-team challenges. At a high level, we find that  existing systems waste effort in {\em irrelevant} tasks unrelated to the challenge, {\em incorrectly execute} tasks, or use {\em brittle post-exploitation techniques}. 
Additionally, all of these systems suffer from significant {\em context bloat} as they proceed  in the complex challenge, which  impacts effectiveness in long-horizon challenges, as seen in other domains~\cite{cemri2025multi}.

Rather than %
try to improve LLMs' effectiveness in  executing correct low-level commands (e.g.,  adding better prompts~\cite{pentestgpt, cyberseceval3}, self-reflection~\cite{penheal, pentestgpt, shao2025craken}), we draw  inspiration from how expert human red teams work and argue that we need to {\em raise the level of abstraction} to build effective autonomous LLM-assisted red teams. 
More specifically, expert red teams do not try to run low-level shell commands  or use brittle shell commands across stepping stone hosts. 
Instead, they think in terms of {\em high-level ``cyber kill-chain'' tasks}~\cite{cyberKillChain} such as  reconnaissance~\cite{oakley2019professional}, exploitation~\cite{mitre_attack}, command and control~\cite{rehberger2020cybersecurity}, and goal-centric actions.
Furthermore, they use best practice {\em  tools} in the security domain to ensure each stage in the kill chain can be effective~\cite{rehberger2020cybersecurity, oakley2019professional, mitre_attack}.

Building on this insight---that raising the level of abstraction is the key---we present the design, implementation, and evaluation of \sys. 
\sys  explicitly {\em decouples red-team planning from execution} (Figure~\ref{fig:intro_before_after}). \sys uses the LLM primarily as a {\em planning} module  that decides \emph{what} task to perform  in terms of {\em high-level declarative tasks}  modeling   cyber kill-chain steps, rather than low-level shell commands. 
\sys  delegates the execution of these tasks to reliable expert task agents using domain-specific best practices. 
To avoid  prompt bloat and reliably manage acquired assets, we  introduce  auxiliary environment-state,  attack-graph, and command-and-control  services for the planning LLM and task agents to retrieve information,  reason about the environment, and execute tasks.  Taken together, these ideas allow \sys to both (a) leverage  the broad world knowledge in the LLM used as a planner to respond and plan next steps and (b) use domain-specific  expertise to execute the plan.

We show that \sys  can handle  unforeseen multi-host scenarios (i.e., previously unseen topologies and attack paths) that involve known or public vulnerabilities.
This setting is representative of many real-world red teams that often chain together known techniques to achieve attack goals in multi-host settings~\cite{equifax_report, darkside, zero_days}. 
The design of \sys is also extensible to include new capabilities (e.g., creating 0-day exploits~\cite{ullah2025cve, wang2025cybergym}).

In some sense, \sys's design represents  a red-team–specific synthesis of best practices in using LLMs for complex and long-horizon tasks---decoupling planning from execution~\cite{yao2023react}, scoped agents~\cite{cemri2025multi}, and offloading solution steps~\cite{gao2023pal, lu2024chameleon, schick2024toolformer}. 
Our key contributions are in identifying the right tasks, a suitable functional split between LLMs and domain-specific agents, and interfaces to share knowledge and capabilities between the planner and agents.

\hyphenation{au-ton-o-mous-ly}

To evaluate \sys, we leverage \benchmark and use
three metrics to capture success: (1) {\em  \acquisition}, indicates whether an attacker has successfully acquired {\em any} critical asset in an environment;  (2) {\em \comprehensiveness}, to measure {\em the proportion} of critical assets captured across multiple  attempts; and (3) {\em \reliability}, to measure the likelihood that any given red-team exercise will succeed at \acquisition.

\begin{figure}[tb]
    \centering
    \includegraphics[width=0.43\textwidth]{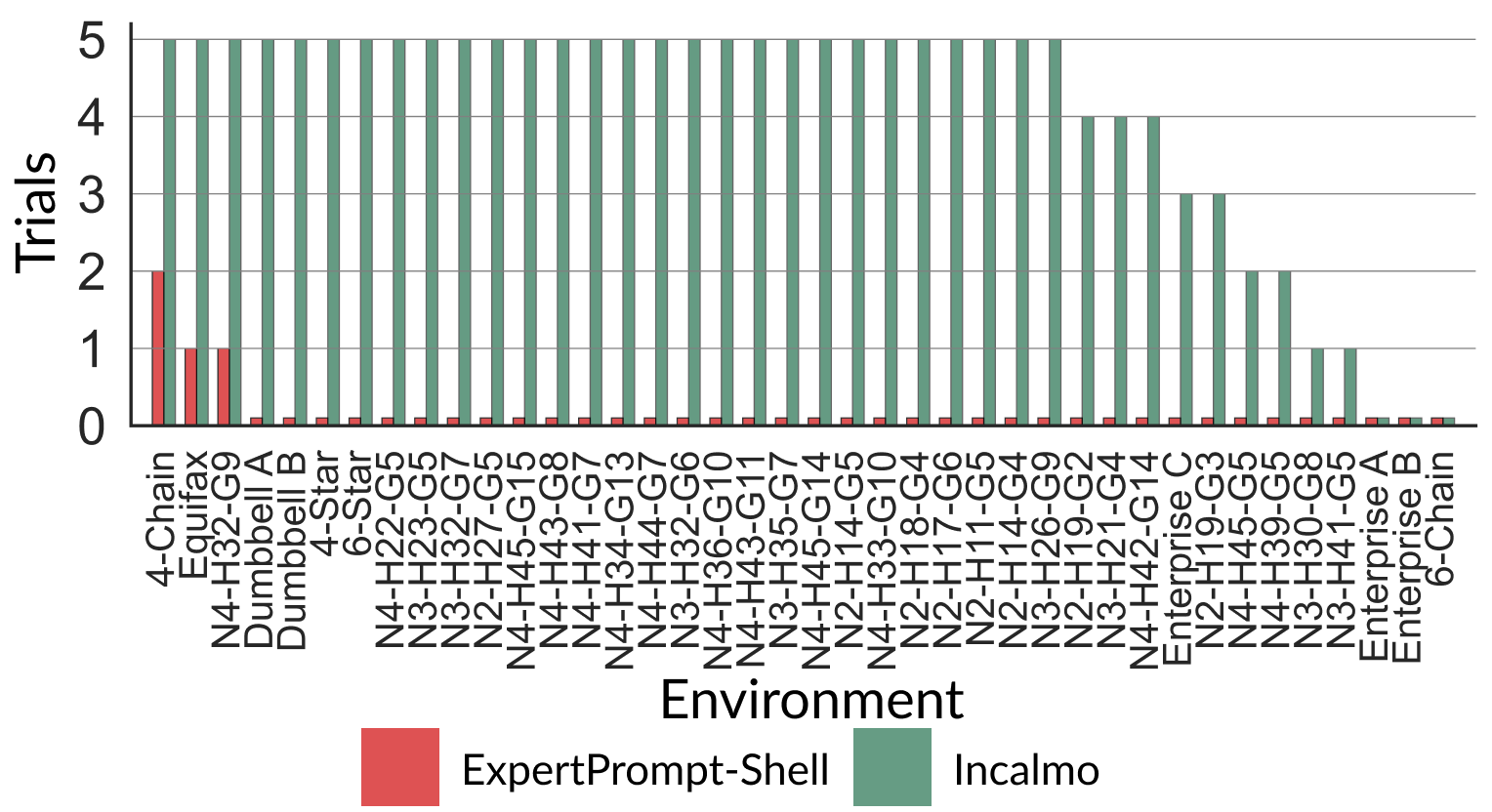}
    \caption{Comparing \reliability across environments between \sys and \sotaSystem with Sonnet 4, the LLM-based system with the best performance on \benchmark.
    \sys succeeded in 37 out of 40 environments while \sotaSystem only succeeded in 3 out of 40 environments.}
    \label{fig:intro_finding}
\end{figure}

As an illustrative finding, \figRef{fig:intro_finding} compares the {\em \acquisition} and {\em \reliability} of \sys against \sotaSystem, the best baseline LLM-based tool\footnote{
\sotaSystem with Sonnet 4 is the best-performing prior system among various baselines, as we show in \secRef{sec:background}.
}, with both systems using Sonnet 4.
In terms of {\em \acquisition}, \sys succeeded in 37 out of 40 environments while \sotaSystem only succeeded in 3. 
In terms of {\em \comprehensiveness}, we find that \sys obtained more critical assets than \sotaSystem in 37 of the environments, with them achieving parity in the remaining 3.  
We find that \sys is cheap and fast---successful attacks took 12--54 minutes and cost $\leq \$15$ in LLM credits.

We  conduct an ablation study to understand which key factors   impact \sys's success.
We show that the choice of LLM used by \sys is not critical and \sys can successfully red-team networks using a variety of LLMs.
Additionally, we find that \sys's abstractions play a larger role than model size: \sys using smaller LLMs (e.g., Haiku 3.5) can   successfully execute attacks in most environments. 

\para{Contributions}
\begin{itemize}

\item  We identify a key gap in existing  cyber offense work and  motivate the need for  LLM-assisted red teaming in multi-host  networks   in unforeseen environments.

\item We develop \benchmark, an extensible benchmark with 40 networks for evaluating LLMs at autonomously executing multi-host red team challenges.

\item We show that leading LLMs with state-of-the-art techniques are largely unable to autonomously execute multi-host red team challenges.
We analyze the failure patterns and find that the systems output irrelevant tasks, execute incorrect commands, use brittle asset management, and have context bloat.

\item We present \sys, building on the idea of  raising the level of abstraction at which the LLM plans,   delegating  execution of high-level tasks to   domain-specific expert agents, and introducing auxiliary services to tackle context bloat and to  manage  acquired assets.   We show that \sys can autonomously obtain critical assets in 37 out of the 40 environments.

\end{itemize}

\para{Ethics and reproducible research}
We acknowledge that \sys is a dual-use technology that could be leveraged by real attackers.
However, systems like \sys can also %
help defenders proactively test their networks.
As prior work has also noted~\cite{zhang2024cybench, pentestgpt}, there is a long history of dual-use systems (e.g., bug finding~\cite{wang2025cybergym, holm2022lore}, exploit generation~\cite{pentestgpt, caldera, metasploit}) helping defenders more than attackers~\cite{silic2013dual} and we believe the same will be true for autonomous red teams.
Additionally, the use of LLMs by real attackers has already been documented~\cite{anthropic2025threat, anthropic2025ai_espionage}, and we hope our work will help defenders also benefit from the capabilities that LLMs offer.
Following the precedent of previous work~\cite{pentestgpt, zhang2024cybench, caldera, xu2024autoattacker, Hu2020AutoPentestDRL, metasploit, kouremetis2025occult, wang2025cybergym}, we are open-sourcing \sys, the benchmarks, and all of the code related to this study to enable defenders and researchers to use and build on our findings. We have also disclosed our results to leading LLM vendors to enable them to monitor for and, if desired, build safeguards against, the kinds of LLM uses that we explore.
We discuss research ethics in more detail in the \hyperref[sec:ethics]{Ethics considerations} section.

\section{Motivation and background}
\label{sec:background}

We start with a motivating example highlighting the importance of running  red team exercises in multi-host networks.
Then, we give a brief overview of related work in cyber-offense  systems.
We address a key gap in prior work~\cite{offsec_sok}---understanding how existing LLM-based systems perform in  {\em multi-host red team exercises}.

\begin{figure}[tb]
    \centering
    \includegraphics[width=0.48\textwidth]{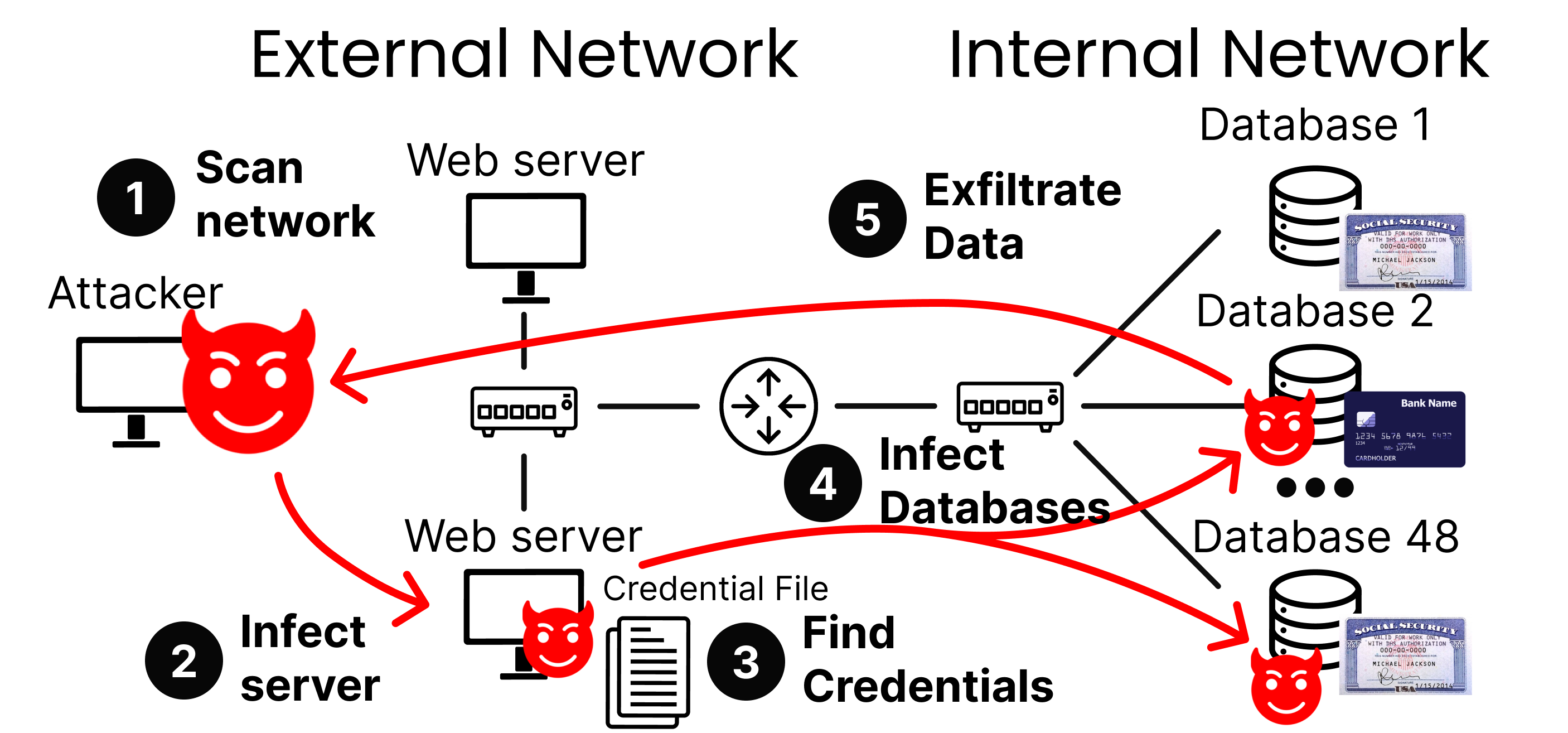}
    \caption{The attacker during 2017 Equifax breach executed a multi-host attack. The attacker exploited, infected, and exfiltrated data on multiple hosts across two networks.}
    \label{fig:equifax_example}
\end{figure}

\subsection{Motivating example: Red teaming Equifax}
We begin by highlighting the importance of red teams in multi-host settings using  the 2017 Equifax data breach as an illustrative example~\cite{equifax_report}, shown in \figRef{fig:equifax_example}.
Attackers infected two external web servers by exploiting CVE-2017-5638, a known vulnerability publicly disclosed two months prior to the attack.
Attackers then found plain text credentials on one of the web servers, used the credentials to compromise database servers, and then exfiltrated  sensitive user data from 48 databases. 
This example illustrates the  {\em multi-host} and ``stepping stones'' nature of a real-world attack spanning multiple network segments and different vulnerabilities to acquire the critical asset(s). 
 
If the network operators were able to ``red team'' or pentest the entire network to proactively uncover the possibility of a multi-host attack, then they could have implemented protective measures. 
For instance, it  may have flagged that data exfiltration monitoring rules were not being actively monitored~\cite{equifax_ftc_settlment}.
Or it may have highlighted how unpatched vulnerabilities and plaintext credentials could be chained together to exfiltrate critical data~\cite{equifax_ftc_settlment}, to help prioritize these problems to operators.

Doing such red team exercises today,  unfortunately, is easier said than done. 
Such exercises require manual effort from specialized and expensive  teams of experts~\cite{rehberger2020cybersecurity}.

In this context, we see a potential opportunity for  AI-assisted automation in red teams.
Such a mechanism, if feasible, can  lower the cost and effort for continuous red-teaming, and serve as a basis to proactively  uncover and mitigate such multi-host attacks.

\subsection{Approaches to offense systems}
We  categorize existing cyber-offense approaches along three dimensions: 
(1) type of attack challenge (e.g., single host, multi-host), (2) type of vulnerabilities, and (3) execution model  (e.g., LLM vs.\  non-LLM).
Our  focus is on red teaming involving {\em multi-host} challenges executed {\em autonomously}  by {\em LLM-based} systems~\cite{offsec_sok}.

\para{Type of attack challenge} Prior studies evaluate LLMs for  solving CTF-style challenges~\cite{pentestgpt, cyberseceval3,  o1systemcard, zhang2024cybench, anthropic_AISI, penheal, happe2023getting, yang2023language, shao2024nyu_ctf, intercode_ctf, anurin2024catastrophic, fang2024teams, shao2024empirical}.
Many problems do not involve infecting a host (e.g., solving a cryptography challenge~\cite{zhang2024cybench, pentestgpt, o1systemcard}).
Some challenges involve a single action to infect a single host~\cite{happe2023getting, yang2023language, intercode_ctf, fang2024teams, shao2024empirical}.
More difficult challenges are single-host attacks that involve completing a series of steps~\cite{pentestgpt, zhang2024cybench, shao2024nyu_ctf, cyberseceval3,  o1systemcard, anthropic_AISI, happe2025can}.
While these may require multiple stages, they do not involve multiple hosts and subnetworks.

We refer to  challenges that involve multiple hosts and subnetworks as {\em multi-host network attacks}.
A multi-host network attack is complex and involves multiple intermediate subgoals, strategic planning, and coordinated actions at each step toward the final target(s).

\newcolumntype{Y}{>{\raggedright\arraybackslash}X}  %
\newcolumntype{C}{>{\centering\arraybackslash}X}    %

\newcommand{\abbrcite}[2]{%
  \mbox{#1\cite{#2}}\xspace}

\newcommand{\TBitemsep}{}

\begin{table}[tb]
  \caption{Summary of existing cyber-offense tools and their evaluation environment}
  \label{tab:attack_taxonomy}

  {\scriptsize %
  \setlength{\tabcolsep}{1pt} %
  \renewcommand\arraystretch{0.9}             %
  \centering
  \begin{tabularx}{\columnwidth}{@{}YCCCC@{}}
    \toprule
    \textbf{Evaluation environment}
      & \multicolumn{2}{c}{\textbf{Non‑LLM}} 
      & \multicolumn{2}{c}{\textbf{LLM}} \\ 
    \cmidrule(lr){2-3}\cmidrule(lr){4-5}
      & Manual & Automated & Semi‑auto & Automated \\ 
    \midrule

    Single‑stage Single‑host \includegraphics[width=1.5cm]{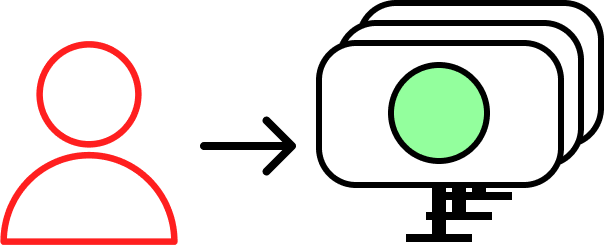}&
      \abbrcite{MSF}{metasploit}\TBitemsep
      \abbrcite{NM}{nmap}\TBitemsep
      \abbrcite{HC}{hashcat} & – &
      \abbrcite{PT}{pentestgpt}\TBitemsep
      \abbrcite{CB}{zhang2024cybench} &
      \abbrcite{GP}{happe2023getting}\TBitemsep
      \abbrcite{YL}{yang2023language}\TBitemsep
      \abbrcite{IC}{intercode_ctf}\TBitemsep
      \abbrcite{FT}{fang2024teams}\TBitemsep
      \abbrcite{SH}{shao2024empirical} \\ 

    \midrule
    Multistage Single‑host \includegraphics[width=2cm]{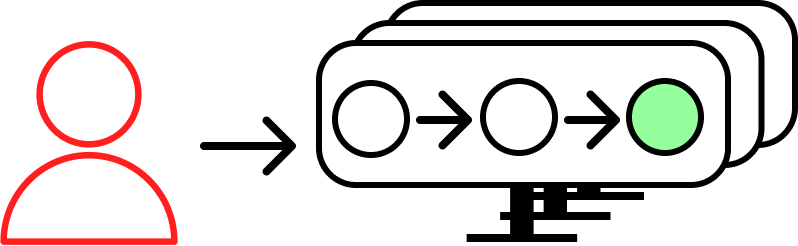} &
      \abbrcite{MSF}{metasploit} &
      \abbrcite{CD}{caldera} &
      \abbrcite{PT}{pentestgpt}\TBitemsep
      \abbrcite{CB}{zhang2024cybench}\TBitemsep
      \abbrcite{AA}{xu2024autoattacker}\TBitemsep
      \abbrcite{AP}{al2025pentest} &
      \abbrcite{NY}{shao2024nyu_ctf}\TBitemsep
      \abbrcite{CA}{mayoral2025cai}\TBitemsep
      \abbrcite{CS}{cyberseceval3}\TBitemsep
      \abbrcite{O1}{o1systemcard}\TBitemsep
      \abbrcite{CB}{zhang2024cybench}\TBitemsep
      \abbrcite{AT}{anthropic_AISI}\TBitemsep
      \abbrcite{VB}{kong2025vulnbot} \\ 

    \midrule
    Multistage Multi‑host \includegraphics[width=3cm]{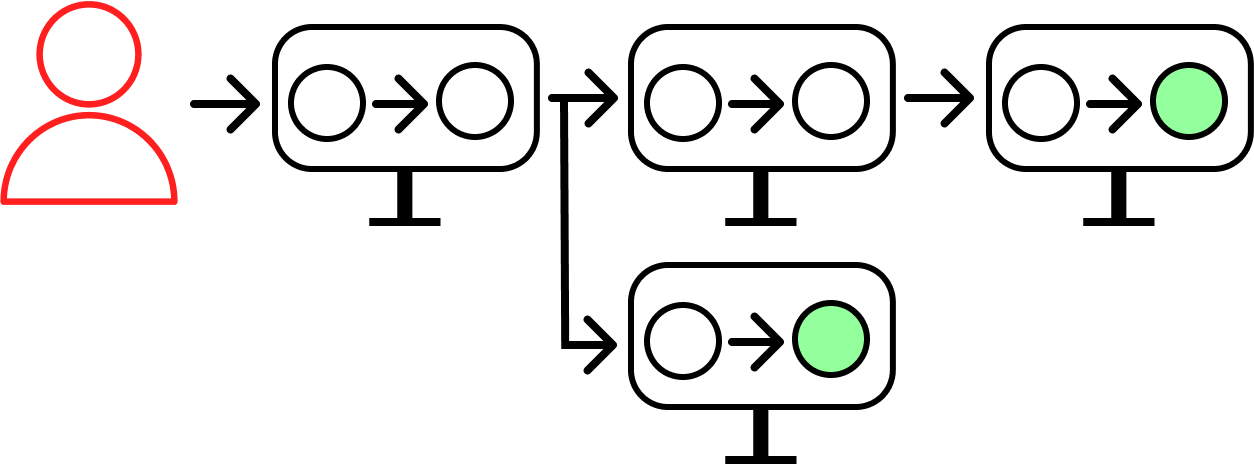} &
      \abbrcite{MSF}{metasploit} &
      \abbrcite{CD}{caldera}\TBitemsep
      \abbrcite{LR}{holm2022lore}\TBitemsep
      \abbrcite{HR}{enoch2020harmer}\TBitemsep
      \abbrcite{CY}{yoo2020cyber}\TBitemsep
      \abbrcite{AJ}{ajmal2021offensive}\TBitemsep
      \abbrcite{SV}{sved_tool}\TBitemsep
      \abbrcite{HP}{Hu2020AutoPentestDRL}\TBitemsep
      \abbrcite{DE}{DeepExploit} &
      – &
      \sys\ \newline
      \abbrcite{OC}{kouremetis2025occult}\footnotemark\TBitemsep
      \\
    \midrule
     & Legend  &  \includegraphics[width=3cm]{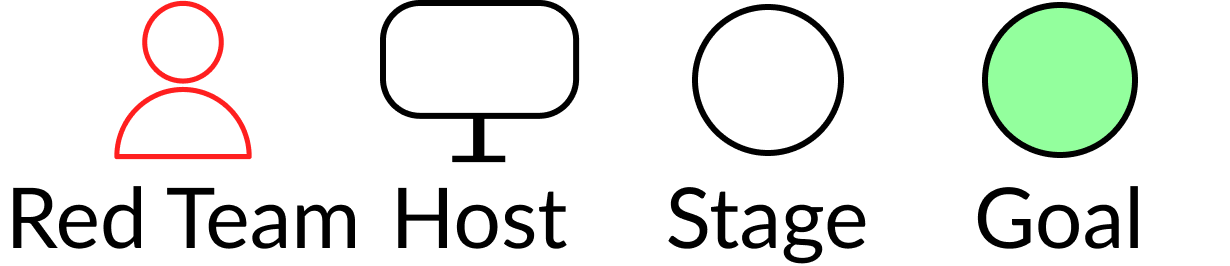} &   
  \end{tabularx}
  }%
\end{table}

\footnotetext{The MITRE OCCULT is a framework to understand the cyber security risks of LLMs.
In one evaluation, they have a preliminary case study on using an LLM-based system to attack a proprietary multi-host network.
Later, we evaluate \sysNoActions, a similar approach to the case study, on \benchmark and show that LLMs fail to even partially succeed in any environment.}

\para{Type of vulnerabilities} Some prior offense systems  assume vulnerabilities are known~\cite{cyberseceval3, zhang2024cybench, pentestgpt, caldera}, while others focus on 0-day vulnerabilities~\cite{wang2025cybergym, shao2025craken}.
As a first step in the multi-host setting, we focus on attacks using known vulnerabilities which is a serious concern in  real-world attacks~\cite{equifax_report, colonial_pipeline_techtarget, zero_days}.

We summarize prior work as seen in Table \ref{tab:attack_taxonomy}: (1) LLM-based systems and (2) Non-LLM-based systems.

\para{LLM-based systems} LLM-based autonomous offense systems~\cite{happe2023getting, yang2023language, intercode_ctf, fang2024teams, shao2024empirical, shao2024nyu_ctf, cyberseceval3,  o1systemcard, zhang2024cybench, anthropic_AISI, mayoral2025cai} entail instructing LLMs to attack the environment. 
The LLM outputs shell commands, which a second entity (e.g., human, MCP server) executes on a computer with access to the environment, shown in \figRef{fig:intro_before_after}.
The output of the command is optionally processed~\cite{pentestgpt, mayoral2025cai} and appended to the context.
Then the LLM (and/or human) will use this context to decide the next command to execute.

\para{Non-LLM-based systems} There is also work on multi-host attacks that do not rely on LLMs—some fully autonomous~\cite{caldera, holm2022lore, enoch2020harmer, yoo2020cyber, sved_tool}. 
There are rule-based and state-machine-based systems (e.g.,~\cite{holm2022lore, sved_tool, enoch2020harmer, yoo2020cyber, ajmal2021offensive}). 
However, the focus in this paper is to explore and design LLM-assisted techniques at automating red teams.

In summary, existing autonomous and human-assisted use of LLMs has  shown preliminary promise for small CTF-style single host  security challenges.
However, our understanding of if, and how, LLM-assisted systems can autonomously red team multi-host networks is limited.

\subsection{Existing LLM-based systems are ineffective in multi-host red team challenges}
To address the gap in evaluating state-of-the-art LLM-based systems,  we  create a novel  {\em multi-host network attack} benchmark called  \benchmark. We describe \benchmark  in more detail in \secRef{sec:implementation} and \appRef{sec:appendix_environments}.
 In this section, we select  10 illustrative multi-host attack challenges from \benchmark and evaluate the red team effectiveness of the aforementioned baseline systems.

\para{Success criteria}
In real-world multi-host environments, there  are often multiple key assets (e.g., Equifax had multiple sensitive databases in \figRef{fig:equifax_example}).
Similar to human red teams~\cite{oakley2019professional}, we consider an attack successful if an attacker is able to compromise at least one key asset (e.g., exfiltrated SSNs from at least one database).
For the experiments in this section, we consider two kinds of  metrics: {\em  \acquisition} indicates if {\em any} critical asset was acquired;  and {\em \comprehensiveness} to measure {\em how many} critical assets were captured (formal definitions in \secRef{sec:eval}).  

\para{Baselines}
In terms of  autonomous LLM systems, we consider three baselines: (1) CyberSecEval3~\cite{cyberseceval3}, (2) \sotaSystem, a shell system with a prompt we created in collaboration with a domain expert at a leading LLM provider, and (3) CAI~\cite{mayoral2025cai}, a popular open-source system.\footnote{
All prompts are in our open-source repository.
}
We choose these systems because they use a variety of the techniques (e.g., chain-of-thought~\cite{cyberseceval3, wei2022chain}, ReAct~\cite{yao2023react}, and self-reflection~\cite{cyberseceval3, pentestgpt}) and were reproducible with open-source prompts and systems.
For the semi-autonomous or human-in-the-loop systems,  we evaluate PentestGPT~\cite{pentestgpt} because it encompasses many ``reasoning'' strategies in other work (e.g., ~\cite{pentestgpt, zhang2024cybench}). 
Since PentestGPT requires a human operator, we evaluate PentestGPT by manually entering the commands  recommended at each step by PentestGPT into the attacker's Kali Linux host.
For \sotaSystem and CyberSecEval3, we consider 3 state-of-the-art LLMs: Sonnet 4, GPT 4o, Gemini 2.5 Pro.\footnote{We were unable to evaluate OpenAI's ``o'' or ``GPT-5'' models because the public API has a safeguard that prevents them from executing attacks.
}

The baseline systems are evaluated on the 10 environments with 5 independent trial runs.\footnote{Since PentestGPT requires manual effort, we only execute 3 trials with GPT4o, the recommended LLM.}
We  also evaluate a SOTA non-LLM attack system, MITRE's Caldera~\cite{caldera}, a popular open-source tool  for red teaming multi-host networks.
Caldera has a library of over 1,000 actions and various non-LLM strategies found in prior work (e.g., RL~\cite{mirage}, weighted decisions~\cite{caldera2024bountyhunter}).
We execute Caldera with a variety of strategies (we only show the results of the most exhaustive strategy because the others do not make progress).

\begin{figure}[tb]
    \centering
    \includegraphics[width=0.48\textwidth]{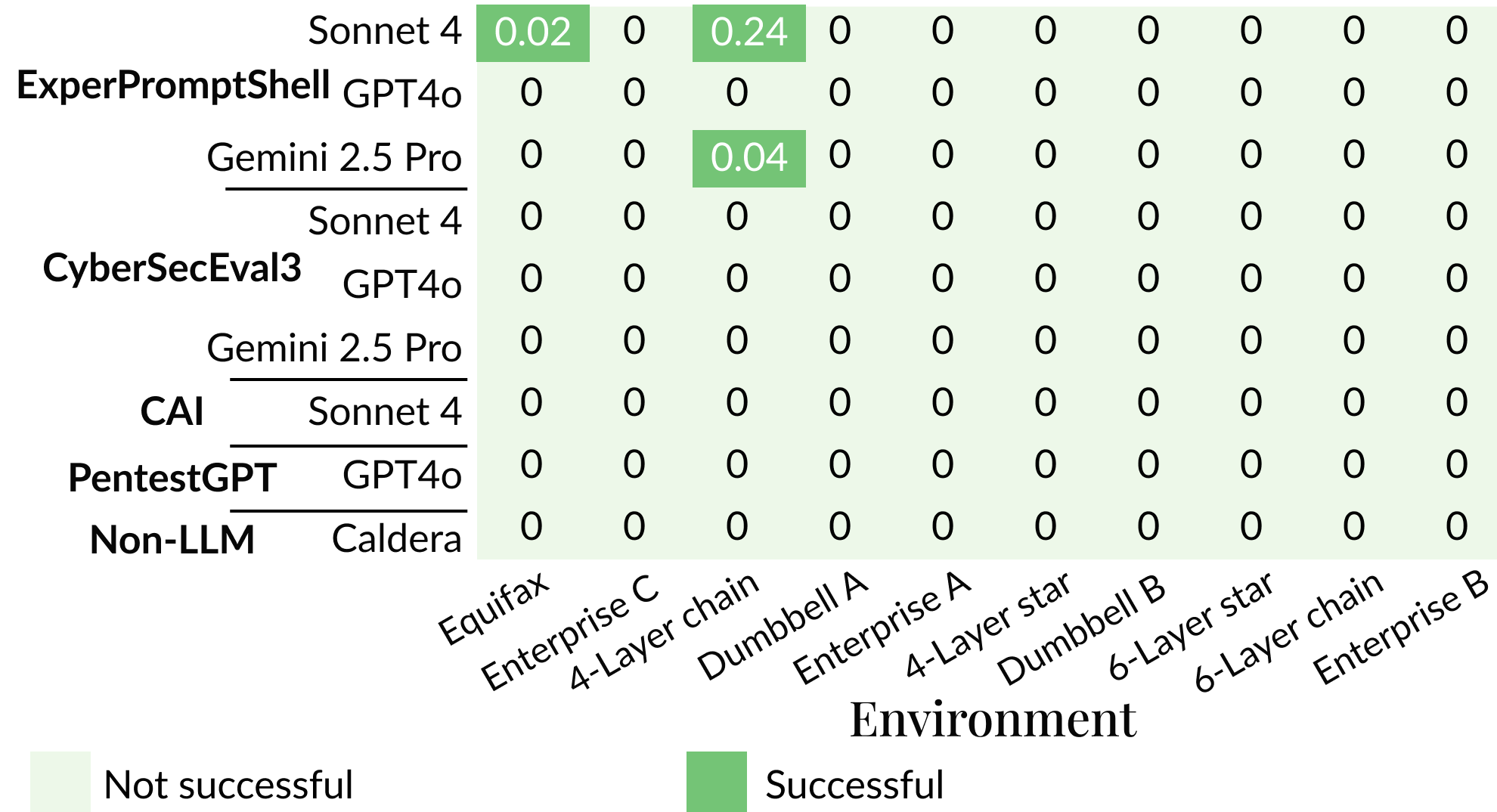}
    \tightcaption{The \acquisition and \comprehensiveness metrics of LLM offense systems across 10 environments. 
    The systems were largely unable to realize multi-host attacks. 
    } 
    \label{fig:bash_goal_summary}
\end{figure}

\para{Findings} 
Across all evaluated LLMs and environments, we find that existing state-of-the-art   LLM-assisted and non-LLM-based systems are  largely unable to realize multi-host attacks w.r.t both \acquisition and \comprehensiveness, shown in \figRef{fig:bash_goal_summary}.
 Only \sotaSystem with Sonnet 4 was able to succeed even partially, by managing to  exfiltrate data from  one of the  database servers in the Equifax-inspired environment.
\sotaSystem with Gemini 2.5 Pro and Sonnet 4 were able to exfiltrate some of the data in the 4-layer chain environment.
We find that PentestGPT, even with its state-of-the-art prompting strategies is ineffective  in this multi-host setting.\footnote{We have tested other models such as DeepSeek and Llama 3 but do not show these results for brevity. 
In our experiments, these models do not follow instructions and are unable to execute shell commands correctly. 
As models get released, we plan to update our benchmark ``scorecard'' (\figRef{fig:bash_goal_summary}).}

\section{Failure analysis }
\label{sec:why_llms_bad}
In this section, we analyze {\em how} existing  state-of-the-art LLM-assisted systems fail at multi-host red team exercises. These insights help  inform our design of \sys.

\subsection{Methodology}
\begin{figure}[tb]
    \centering
    \includegraphics[width=0.48\textwidth]{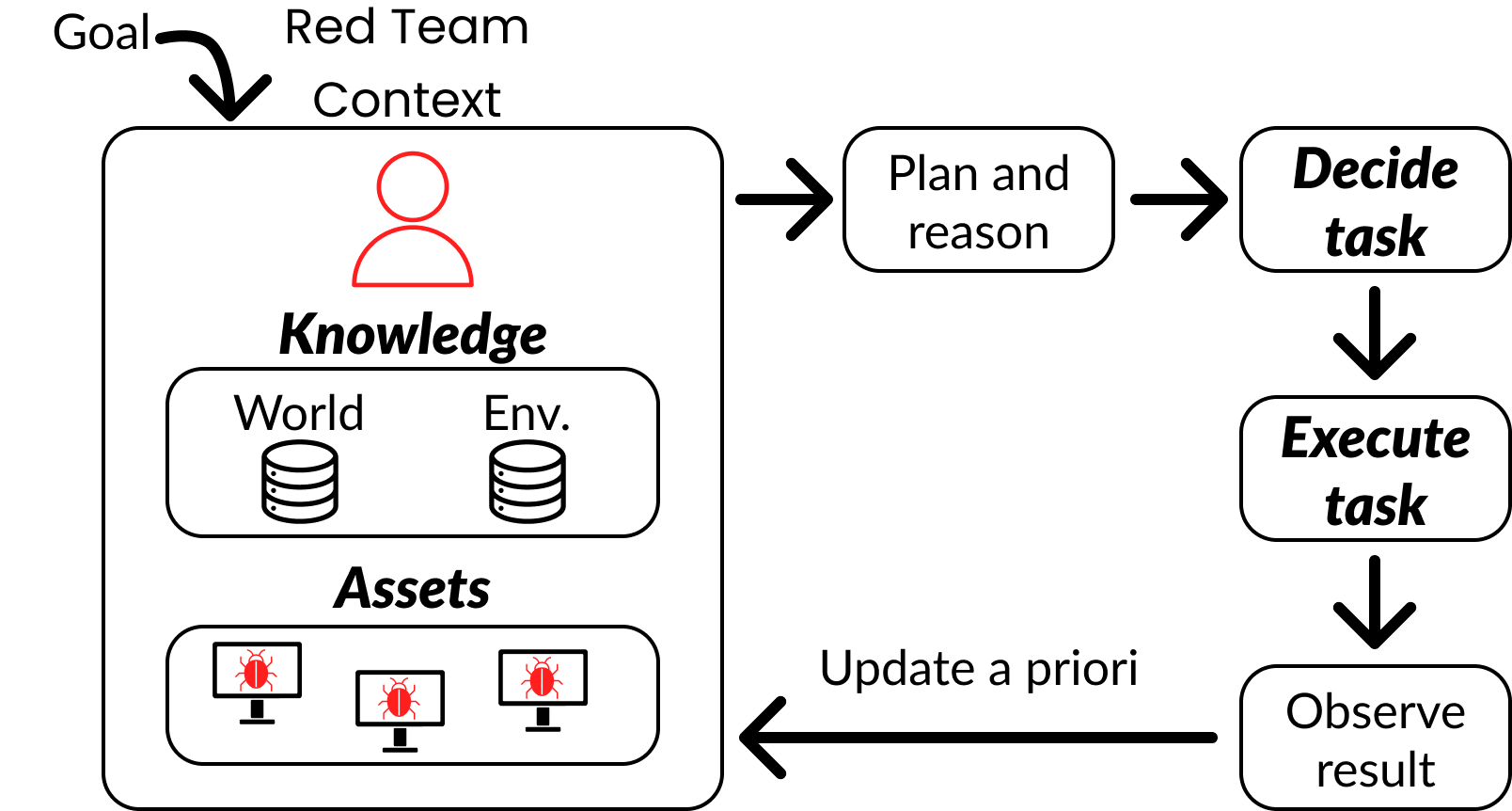}
    \tightcaption{A mental model of how red teams execute multi-host attacks. Red teams start with a goal (e.g., exfiltrate important data). Then, they follow a loop of deciding a task (e.g., infect a server), executing the task (e.g., launching an exploit), and updating their knowledge/capabilities.} 
    \label{fig:mental_model}
\end{figure}

 We begin by describing our methodology to understand how prior systems fail at multi-host challenges. Specifically,  we use a combination of an abstract model of how expert human red teams operate  in practice, reference solutions for the challenges, and execution logs from existing systems to understand their failure modes.
We describe each   next.

\para{Abstract mental model}
Red-team operations operate through an iterative loop shown in \figRef{fig:mental_model}.
Starting from their initial    {\em knowledge} (e.g., known vulnerabilities) and what {\em assets} they can control (e.g., command execution on a host), the red team {\em plans and decides} a next logical {\em task} to implement (e.g., infect a host) ~\cite{rehberger2020cybersecurity}.
The red team will then attempt to {\em execute} the task (e.g., launch an exploit).
A successful task either obtains new assets (e.g., access to a new host) or discovers new {\em knowledge} (e.g., finding sensitive data).
Then, the red team updates their knowledge/asset base and decides the next task.
The red team repeats this loop until they either achieve all goals or runs out of time~\cite{oakley2019professional}.

\para{Reference solutions}
With this mental model, we create reference solutions for how a red team would successfully attack the environments in \secRef{sec:background}.
For each environment, we create a reference solution based on an attack graph model~\cite{attack_graph_og}.
We define a {\em task} as a sequence of commands to reach a state in the attack graph (e.g., found the correct vulnerability, gained access to a server).
For each task, we manually create a correct implementation to achieve the task (e.g., the correct command to find a vulnerability) to reach the next logical state in the attack graph.
For completeness, we provide the details of the attack graph model   in \appRef{sec:appendix_log_analysis}. 

\para{Log analysis}
Using the reference solutions, we manually analyze the logs from the execution runs of the baseline systems from \secRef{sec:background}.  
 This helps us shed light on  common failure modes.    
Given the manual effort of this analysis, we do a qualitative inspection  across  baselines and a deeper dive on the best performing system, \sotaSystem. 
For  \sotaSystem, we first categorize tasks by the LLM-based systems as either relevant or irrelevant.
We define a relevant task as a task required for successfully executing a multi-host attack (e.g., found the correct exploit).
Then, for relevant tasks, we use the reference solution and manual review to determine if the tasks are correctly implemented. %
The details of this analysis are in \appRef{sec:appendix_log_analysis}.

\subsection{Observations}
We now describe our key observations about how existing systems  fail in our  multi-host red team exercises from \secRef{sec:background}. 

\begin{figure}[tb]
    \centering
    \includegraphics[width=0.45\textwidth]{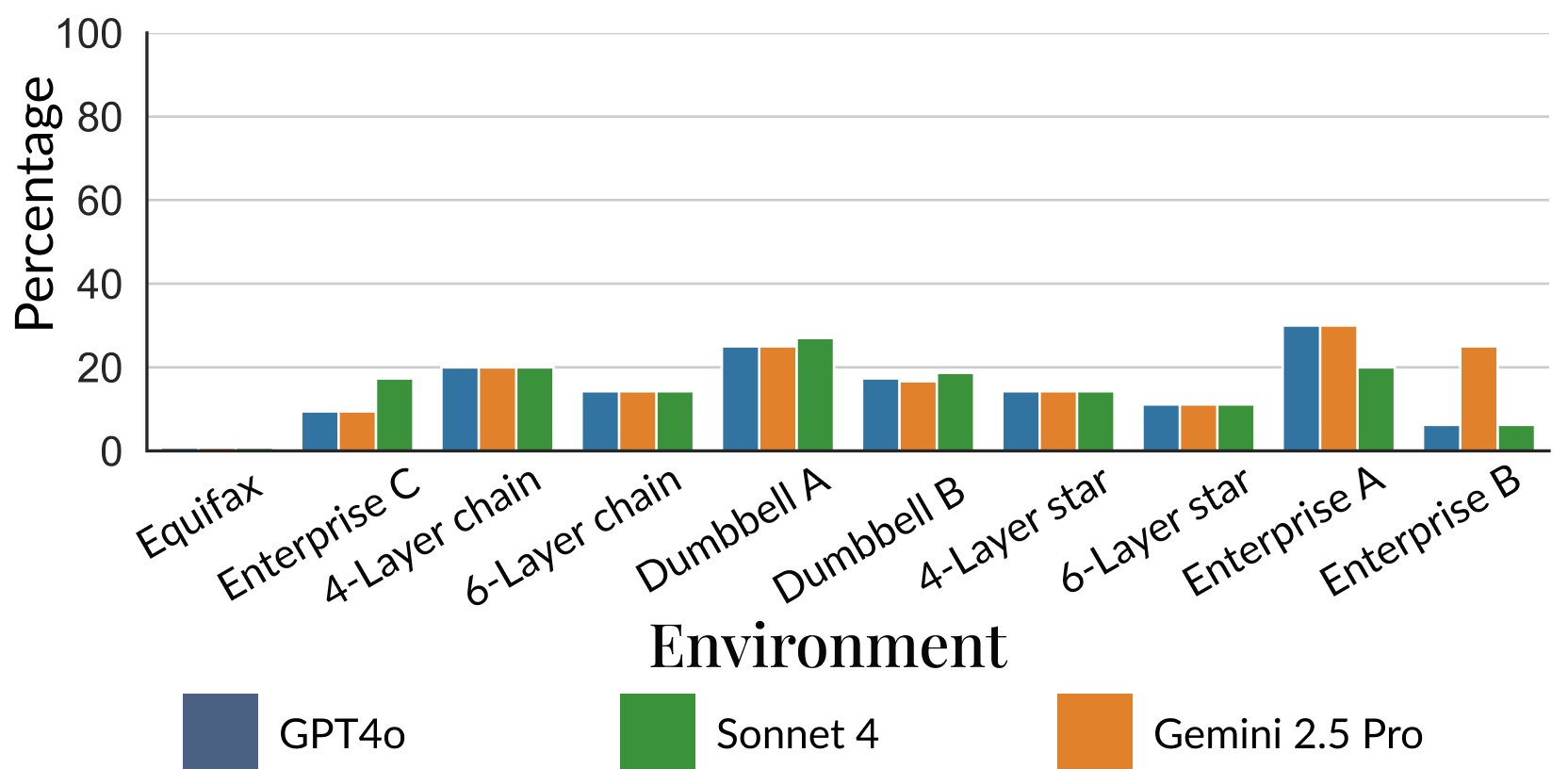}
    \tightcaption{Percentage of tasks successfully implemented by \sotaSystem with different LLMs. Across all environments, \sotaSystem was only able to execute 1--30\% of the tasks.} 
    \label{fig:bash_ag_states}
\end{figure}

\begin{figure}[tb]
    \centering
    \includegraphics[width=0.45\textwidth]{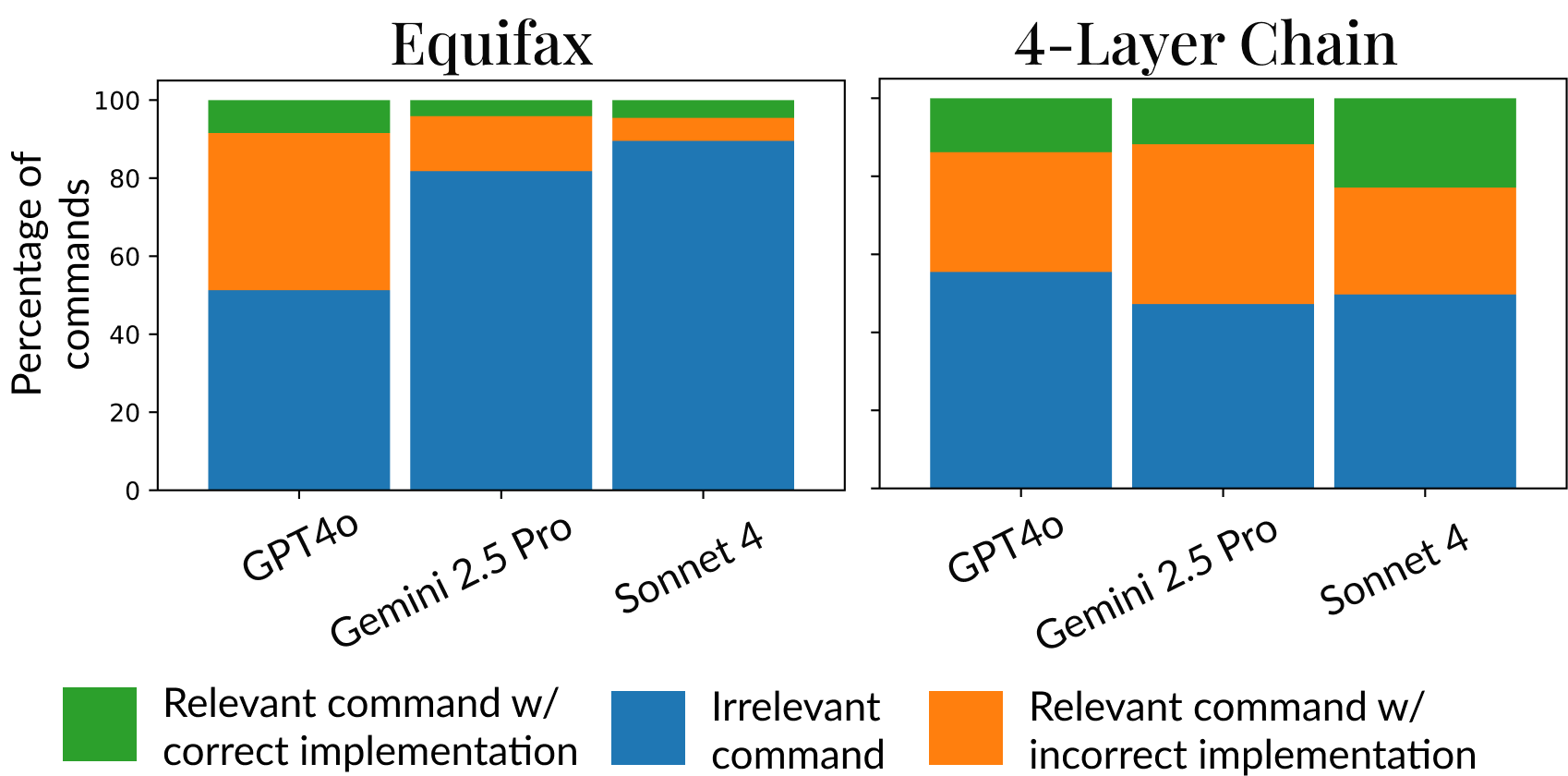}
    \tightcaption{In the Equifax-inspired and chain environments, 47--90\% of \sotaSystem's tasks are irrelevant.
    Furthermore, 6--41\% of \sotaSystem's tasks are implemented incorrectly.} 
    \label{fig:ag_breakdown}
\end{figure}

\textbf{Observation 1: Pursuing irrelevant red team tasks}
We observe that both the LLM-based (and non-LLM-based) red-team systems evaluated in \secRef{sec:background} struggle to correctly decide a task in \figRef{fig:mental_model}.
Across the LLMs and environments, 47--90\% of \sotaSystem's commands are irrelevant, shown in \figRef{fig:ag_breakdown}.
For instance, the \sotaSystem tried brute forcing SSH credentials, finding misconfigured files, or exploiting non-exploitable services.
Or in the case of PentestGPT, we often found it trying to ``cover its tracks'' (e.g., deleting command history) on the attacker's Kali host.
Caldera, a non-LLM  based system  also executed irrelevant tasks; e.g.,  frequently attacking the attacker's  own Kali  host, rather than use the attacker's host for red teaming.

\textbf{Observation 2: Incorrectly executing tasks}
Even when LLM-based systems pursued relevant red teaming tasks, we observed that they struggled to correctly execute these tasks in \figRef{fig:mental_model}.
For instance, across the LLMs and environments, 6--41\% of \sotaSystem's relevant tasks are implemented with incorrect commands (\figRef{fig:ag_breakdown}).
Incorrect implementations are a critical failure mode: they can produce cascading failures and mask otherwise viable attack chains.
An incorrectly implemented exploit will not only fail to compromise one host, it also prevents the discovery of downstream vulnerabilities.

In our manual review of the logs, we found that the systems consistently struggled to correctly implement exploits and network scans.
For example, in one case we see that \sotaSystem with Sonnet 4 tried to create a complex exploit for  the  Apache Struts vulnerability, but the implementation is wrong and fails: 

{\scriptsize
\begin{verbatim}
curl -H "Content-Type: %
(#dm=@ognl.OgnlContext@DEFAULT_MEMBER_ACCESS).
(#_memberAccess?(#_memberAccess=#dm):
((#container=#context['com.opensymphony.xwork2.
ActionContext.container']).
(#ognlUtil=#container.getInstance(@com.opensymphony.
xwork2.ognl.OgnlUtil@class)).
(#ognlUtil.getExcludedPackageNames().clear()).
(#ognlUtil.getExcludedClasses().clear()).
(#context.setMemberAccess(#dm)))).(#cmd='id').
(#iswin=(@java.lang.System@getProperty('os.name').
toLowerCase().contains('win'))).
(#cmds=(#iswin?{'cmd.exe','/c',#cmd}:
{'/bin/bash','-c',#cmd})).
(#p=new java.lang.ProcessBuilder(#cmds)).
(#p.redirectErrorStream(true)).(#process=#p.start()).
(#ros=(@org.apache.struts2.ServletActionContext@
getResponse().getOutputStream())).
(@org.apache.commons.io.IOUtils@copy
(#process.getInputStream(),#ros))%
http://192.168.200.10:8080/showcase.jsp
\end{verbatim}
}

Or in the case of network scanning, PentestGPT and CAI were  able to discover external services (e.g., a web server) through scanning tools such as \texttt{nmap}.
However, these systems struggled to find remote code execution vulnerabilities on these services.
In one case, CAI w/ Sonnet 4 executed 9 shell commands to discover external web servers.
Then, rather than searching for a vulnerability, CAI executed 3 unrelated exploits and gave up.
Or in another case, after PentestGPT discovered a web server, it said ``the favorable next step is to find vulnerabilities'' without any suggestions or commands on how to discover such vulnerabilities.

\textbf{Observation 3: Using brittle post-exploitation techniques to command hosts}
In the few times \sotaSystem successfully executed exploits, the system used brittle post-exploitation techniques to control assets.
We only observe this in \sotaSystem because none of the other systems were able to make substantial progress.
\sotaSystem w/ Sonnet 4 often used exploits to execute commands on vulnerable hosts, rather than establishing an agent connected to a command and control server.
Exploits are often not a reliable method of executing commands~\cite{rehberger2020cybersecurity}, but more importantly the unreliability cascades in multi-host environments.
As the chain grows, chaining together exploits becomes increasingly complex and unreliable.\footnote{In real-world attacks and red team exercises, attackers primarily use exploits to install malware agents that communicate with a C\&C server~\cite{equifax_ftc_settlment, colonial_pipeline_techtarget, rehberger2020cybersecurity}}

We also found that \sotaSystem w/ Sonnet 4 used \texttt{ssh} and reverse shells to execute commands on vulnerable hosts.
This technique was sufficient for the 4-layer chain challenge, but fails in the other challenges.
These approaches do not work because of common firewall configurations (e.g., a web server does not have \texttt{ssh} configured).

\textbf{Observation 4:  Knowledge has context bloat}
All of the prior LLM-systems store all knowledge by adding observations (e.g., command outputs) to the context.
We found this especially in \sotaSystem (the best performing system) and CyberSecEval3, where  the  context grew  with many low-level implementation details clogging the red team's knowledge in \figRef{fig:mental_model}.
For instance, in one case \sotaSystem with Sonnet 4 on the Enterprise A environment executed 108 shell commands with a final context of 54K tokens (157,760 characters).
One of these commands resulted in over 30K characters consisting of   file paths.
These long contexts likely cause  the LLM systems to struggle to maintain a high-level plan~\cite{cemri2025multi, pentestgpt}.\footnote{ 
Interestingly, the authors of PentestGPT~\cite{pentestgpt} noted this same problem when solving CTF challenges and introduced a token compression module to limit the context. 
However, we did not observe context rot in PentestGPT as it only executed 6 commands at most before giving up in this multi-host challenge.}

\section{\sys: An LLM-based system for autonomously executing multi-host red teams}
\label{sec:high_level_api}
In this section, we begin by describing the high-level idea underlying \sys before presenting the detailed design.

\subsection{High-level idea}
Our design of \sys draws from both our mental model of expert red teams and the failure modes we observed in existing LLM-assisted  systems.
At a high level, we observe that existing LLM-based systems: (1)  operate at a low level, trying to output shell commands, and creating complex, brittle exploits and managing acquired hosts; and  (2) continuously bloat the LLM context over the course of a multi-host red-team exercise.

Building on these lessons,  
we argue for an approach that {\em raises the level of abstraction} at which an LLM-assisted red team operates.  To this end, we explicitly {\em decouple planning from execution} by separating \sys into an LLM-assisted planning layer that decides \emph{what} tasks to perform and an execution layer that decides \emph{how} to execute tasks.

\begin{figure}[tb]
    \centering
    \includegraphics[width=\columnwidth]{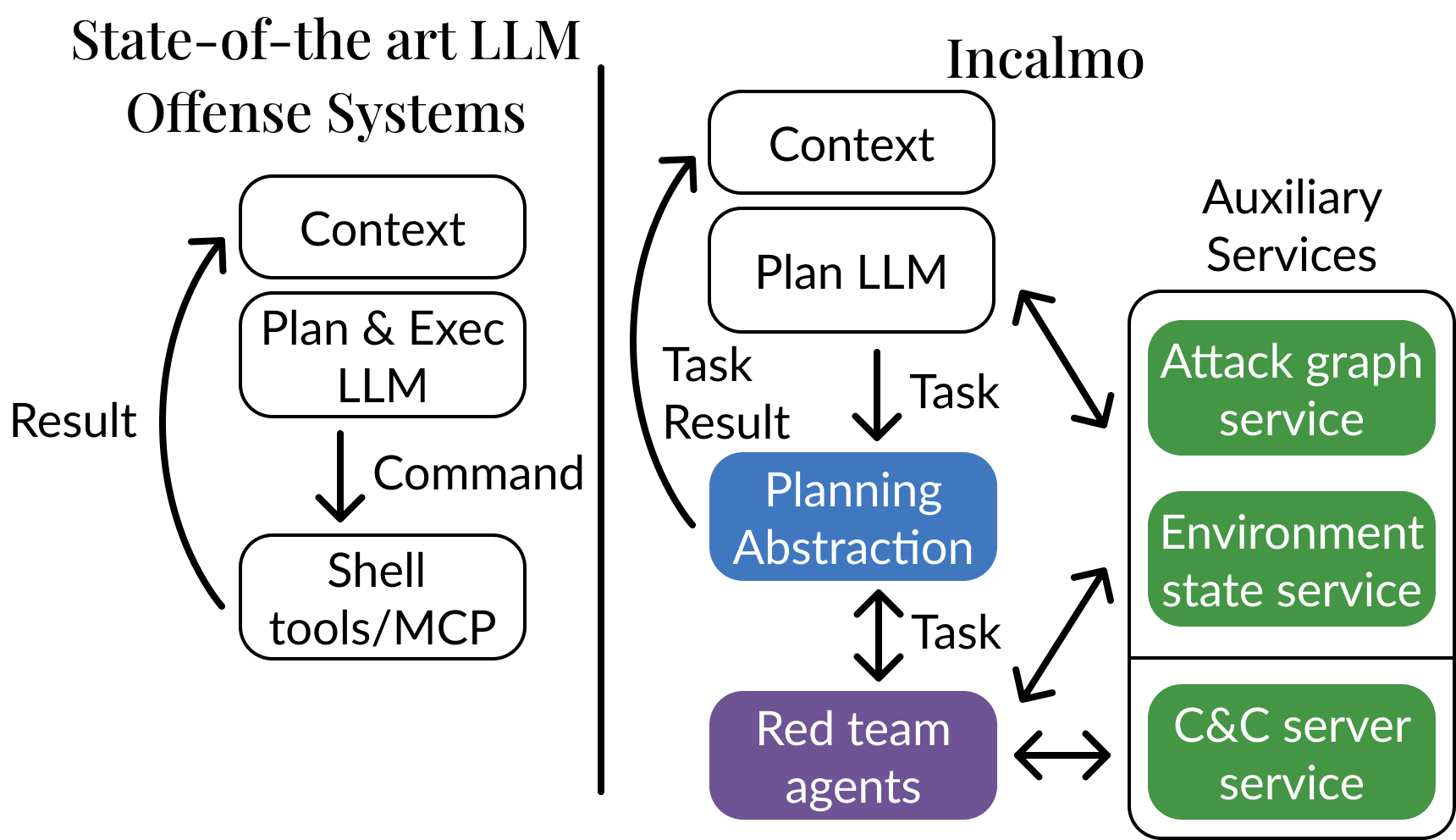}
    \caption{\sys\ uses LLMs to plan multi-host attacks with high-level tasks. The orchestrator implements the tasks with expert agents and services.}
    \label{fig:sys_overview}
\end{figure}

This is in contrast to how prior systems use LLMs to both plan and execute tasks, shown in \figRef{fig:sys_overview}.
Rather than have the planner output low-level commands  as  done in the baseline systems, we reformulate it to output  high-level declarative tasks inspired by the cyber kill chain framework~\cite{cyberKillChain, mitre_attack}. 
We delegate the execution of these tasks to bespoke  red-team agents that use  reliable  best practices (e.g., a C\&C server service). 
This is in contrast to how prior systems used a variety of heuristics  (e.g., fine-tuned system prompts~\cite{pentestgpt, cyberseceval3, zhang2024cybench}, command self-reflection~\cite{pentestgpt,mayoral2025cai, penheal}, summarizers~\cite{pentestgpt, penheal}) to improve the ability of a single LLM to combine planning and execution. 
Finally, to tackle the context bloat, we introduce auxiliary environment-state and attack-graph services that can be queried (akin to   RAG~\cite{lewis2020retrieval}) by the planner and expert agents. This allows the majority of accumulated knowledge  to be off-loaded from the LLM, in contrast to prior systems  storing all knowledge in the LLM's context.

\para{Scope and limitations} We scope our design in this paper with two known limitations. First,
 our  design  does not take into account  defender capabilities (e.g., detection, blocking),   similar to prior work in LLM-based offense evaluations~\cite{pentestgpt, cyberseceval3, zhang2024cybench}. 
Second,  similar to many real-world settings,  we assume the red team exercise only considers known vulnerabilities and  do not consider zero-day exploits~\cite{shao2025craken}. We do note, however, that 
\sys  is extensible and could include these considerations in future work. 

\subsection{Detailed design}
Having described the high-level idea above,  next we describe our concrete realization of the: (1)  planning abstraction and declarative tasks; (2)  library of red team agents to implement the  specific tasks; and (3)  auxiliary services that enable the planner LLM and the task agents to reliably manage knowledge and assets they have gained during the red team exercise.

\para{Planning abstraction} 
Prior systems plan and execute red teams in terms of  low-level shell commands and tools.  
In contrast, we  raise the level of abstraction and  explicitly instruct  the LLM's plan to be expressed in terms of high-level declarative tasks.
More specifically, our abstraction for these declarative tasks    follows the logical stages specified by the  MITRE ATT\&CK~\cite{mitre_attack} and cyber kill chain~\cite{cyberKillChain} frameworks: scan a network, laterally move, escalate privileges, discover local information, and exfiltrate data.
 Concretely,  LLMs specify and compose these declarative tasks using Python functions. 
The functions can use the standard Python library and \sys's API.
 A function can either (1) output a series of high-level tasks (e.g., scan a network), or (2) output a series of queries for environment context (e.g., find hosts on a public network).
 This functional specification  allows the LLM to use the \sys services to specify red team plans.
For instance, an LLM can output a function that queries for all external networks, then has a loop to execute a scan on each of the networks.
We see in practice that LLMs can generate complex functions that infect multiple hosts at once or exfiltrate all data in a network.

\para{Task agents}
We design task agents to translate the above set of declarative   tasks into low-level commands,  as described briefly  in Table \ref{tab:action_translation}. 
Prior LLM-based offense systems  used a variety of techniques to improve command accuracy such as  adding LLM inference methods: self-reflection to correct wrong commands~\cite{pentestgpt, mayoral2025cai, penheal}, increasing the library of low-level MCP tools~\cite{mayoral2025cai}, and even creating a library of system prompts tuned to specific security tasks~\cite{pentestgpt}.
However, in \secRef{sec:background} and \secRef{sec:why_llms_bad}, we observe that these techniques are insufficient in fixing implementation problems for multi-host environments.

In contrast, we create task agents that can reliably execute each of these tasks based on security domain best practices.\footnote{
Later in \secRef{sec:eval}, we also explore the use of LLM-based agents and find that they do not perform well at executing tasks.
} 
For instance, the  lateral movement agent queries the attack graph service to identify possible vulnerable paths to the target server. 
Then, the agent searches an exploit database (e.g.,  Metasploit) for exploits matching these vulnerabilities and executes them. 
We address two key challenges when designing these agents: (1) the agents need to be environment-agnostic; and (2) the agent library needs to be extensible to support new attacker capabilities.

\begin{table}
  \caption{How \sys non-LLM task agents translate tasks}
  \label{tab:action_translation}
  \centering
  \footnotesize         %
  \setlength{\tabcolsep}{2pt}
  \begin{tabularx}{\linewidth}{%
      >{\raggedright\arraybackslash}p{0.33\linewidth} %
      X}                                               %
    \toprule
    \textbf{High‑level tasks} & \textbf{\sys agent translation} \\
    \midrule
    \texttt{\footnotesize FindInformation}&
      Searches common directories for key data and credentials. \\
    \texttt{Scan} &
      Runs \texttt{nmap}/\texttt{nikto} to find vulnerable services. \\
    \texttt{LateralMove} &
      Searches for and executes exploits from an internal library or Metasploit's library.\\
    \texttt{EscalatePrivilege}&
      Searches for and executes exploits from an internal library or Metasploit's library \\
    \texttt{ExfiltrateData} &
      Finds shortest path to attacker's host and then exfiltrates the data.\\
    \bottomrule
  \end{tabularx}
\end{table}

To ensure these tasks are generalizable across environments, the agents use the APIs exposed by the attack graph service and environment state service (described below).
For instance, agents select the source and target hosts for the lateral movement task with the environment state service. 

 We design \sys to be extensible and note  that both the set of abstract tasks and their specific implementations are extensible. 
Since we decouple the task API from the realization, tasks can accommodate multiple execution agents.
For instance, in \secRef{sec:eval_factor_analysis}, we add LLM-based agents as alternative implementations.
We also enable developers to add new high-level tasks.
For instance, users can add a ``stealth data exfiltration'' task (examples in the open-source repository).

\para{Auxiliary services}
Next, we detail the design of the three auxiliary services that help LLMs retrieve relevant context and enable reliable execution of red team tasks: (1) an environment state service to maintain environment knowledge, (2) an attack graph service to reason about potential attack paths, and (3) a C\&C server service to reliably maintain and execute commands on assets.

\emph{(1) Environment state service:} To  tackle  context bloat,  
PentestGPT and CAI use several heuristics such as using LLMs to summarize prior context (e.g., command outputs, chain-of-thought, etc). However,  relevant  information can still get buried in the context: a crucial clue could be discovered on a host, but it only becomes meaningful after several commands on a different host are executed.
While this may not have been a critical flaw in  single-host CTF challenges,  this  stale context quickly becomes a bottleneck in long-horizon, multi-host exercises.

To avoid this context bloat, we design a queryable environment state service that maintains a structured knowledge base of the environment (akin to RAG~\cite{lewis2020retrieval}). 
LLMs can output high-level code   that queries the service.
The idea is that planning LLMs (and agents) query the environment state service for information when it becomes relevant for the attack.
There are two challenges when designing an environment state service: (1) our knowledge of the network changes as attackers run tasks (e.g., a scan discovers a host);  and (2) this knowledge needs to be exposed in a systematic way so the LLM can ``reason'' about the network (e.g., what services does a host have).
To address these challenges, the environment state service maintains a structured database of Python objects that represent the environment, similar to Lore~\cite{holm2022lore}.\footnote{Lore uses traditional state-space exploration tools and algorithms for attack exploration, and is not designed to be exposed to LLMs as such.} 
The database is updated as red team agents execute tasks.
For instance, if an agent discovers hosts with a scan, the database will update and contain objects representing the new hosts.

\emph{(2) Attack graph service:}
We also introduce an attack graph service that helps \sys's planning LLMs and agents correctly decide what tasks to execute~\cite{attack_graph_og}. 
 Multi-host environments are complex and it  is difficult for LLMs to  reason about how vulnerabilities can chain together, especially since attackers have {\em incomplete and evolving} information.
We design a dynamic attack graph service to help both the planning LLM and the agents reason about environments.
 Existing attack graph tools are often developed from a static defense perspective and assume complete and prior knowledge of the network~\cite{OG_MULVAL, ou2006scalable}. 
Our attack graph service  dynamically  retrieves  the current best knowledge   from the environment state service and can recommend the next best course(s) of action for the red team exercise.  
For instance, \sys's lateral move agent queries the attack graph service to identify tasks to infect a host with the following query:
\begin{lstlisting}[language=Python, 
caption={}]
attack_graph_service.get_possible_attack_paths(target_host)
\end{lstlisting}

\noindent When calling this endpoint, the attack graph service will query the environment state service to obtain the current world view about  host vulnerabilities and host reachability criteria to suggest next  hosts to exploit.
 Our current implementation uses a simple  brute-force search to discover   these candidate  paths, which we have found to be sufficiently scalable  for environments on the order of 100s of nodes.

\emph{(3) C\&C server service}
We design a high-level C\&C server service to help task agents reliably execute commands and manage their assets.
Instead of using low-level shell commands to manage assets, we abstract a C\&C server as a service that (A) executes commands on an arbitrary infected user on a host, and (B) has an API endpoint to download and execute malware to infect additional hosts.
Our current implementation   handles all low-level communication techniques (e.g., proxying~\cite{metasploit}, beaconing~\cite{caldera}) internally.
However, the C\&C server service API can be extended to include options to configure these.

\subsection{Illustrative case study}
\label{sec:usage_examples}
\begin{figure}[tb]
    \centering
    \includegraphics[width=0.48\textwidth]{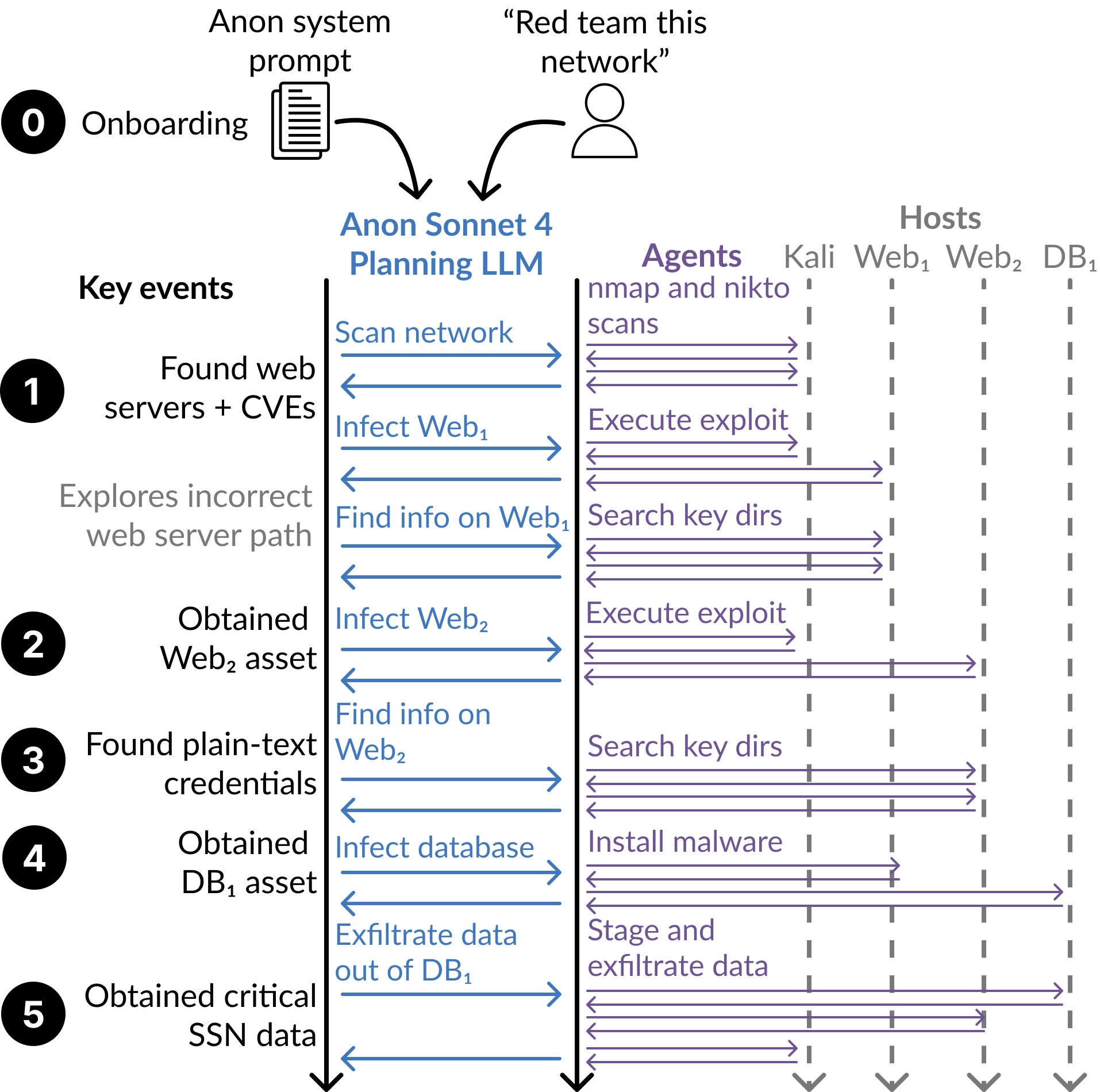}
    \tightcaption{A timeline of \sys red teaming the Equifax environment with Sonnet 4. 
    The red team stages correspond with the stages of the real Equifax attack show in \figRef{fig:equifax_example}.
    \sys uses Sonnet 4 to plan several high-level red team tasks which are executed across several hosts by \sys's agents.} 
    \label{fig:equifax_timeline}
\end{figure}

In this section, we show a concrete end-to-end example of \sys using Sonnet 4 running in an interactive loop to run a red team exercise in  the Equifax-inspired environment.
In this example, \sys w/ Sonnet 4 is following similar steps as the real Equifax attacker in \figRef{fig:equifax_example}.
In \figRef{fig:equifax_timeline}, we both show a timeline of the steps \sys took to red team the network and map key events to the real attack in \figRef{fig:equifax_example}.
The full prompt and logs of this case study are in our open-source repository (see \hyperref[app:open_science]{Open science section}).

\para{Onboarding}
First, we have an LLM-agnostic {\em system prompt} stage where we teach the planning LLM the available capabilities and APIs in \sys.
We also provide the user's {\em environment specific} prompt to outline attack goals and environment details (e.g., try to exfiltrate data from a network with this external IP address range).

\para{Execution}
\sys then uses a Sonnet-4-assisted workflow to plan and execute the red team exercise, shown in \figRef{fig:equifax_timeline}.
Sonnet 4 outputs tasks, \sys agents execute them, the agents return any results or errors, and then Sonnet 4 reacts and decides the subsequent  task(s), as shown in \figRef{fig:equifax_timeline}.

Sonnet 4 first decided to scan Equifax's external network.
The \sys scanning agent   discovered   web servers and identified  that  they have remote code execution  vulnerabilities (\circled{1}, \figRef{fig:equifax_timeline}).
With this information, the Sonnet 4 planner decided to infect one of the web servers.
\sys was able to infect the web server (through the lateral movement agent executing an exploit and installing malware).  However, this  turned out to be a dead end because \sys cannot use this  web server to obtain further network access.

After exploring this futile path, the Sonnet 4 planner decides to infect the other web server (\circled{2}, \figRef{fig:equifax_timeline}).
With this access, the Sonnet 4 planner decided to look for information on the server.
The find information agent used the C\&C server connection to reliably execute commands and found plain-text \texttt{SSH} credentials (\circled{3}, \figRef{fig:equifax_timeline}).
With these credentials, Sonnet 4 decided to infect all of the databases, again using  the lateral movement agent (\circled{4}, \figRef{fig:equifax_timeline}).
Finally, Sonnet 4 decided to exfiltrate the data from the database.
The data exfiltration agent used the environment and attack graph services to identify an exfiltration path out of the network: copy the data to a web server, and then download the data to the attacker's computer over \texttt{HTTP} (\circled{5}, \figRef{fig:equifax_timeline}).
\sys then proceeds to use this workflow in a loop to infect and exfiltrate data from each of the 48 databases in the network (not shown in  \figRef{fig:equifax_timeline} for brevity).

\section{Implementation}
\label{sec:implementation}
We implement \sys as a Python framework consisting of around  8K lines of code.
We implement a custom C\&C server and use open-source malware  capabilities (from the Caldera project~\cite{caldera}) to infect and send shell commands to hosts.\footnote{We picked Caldera given our familiarity.
Other C\&C servers such as  Cobalt Strike~\cite{cobalt} or Merlin~\cite{merlin} could also be used.}
We implement \sys with custom Python modules.
For the environment state service, we create parsers that interpret command outputs and update the knowledge base.

For each of the five high-level tasks in \secRef{sec:high_level_api}, we create non-LLM and LLM-based agents that translate the tasks into low-level primitives (e.g., Python scripts, shell commands).
We implement the non-LLM agents for the lateral movement and privilege escalation tasks by integrating into \sys's internal library (or optionally Metasploit's library) of known vulnerabilities and their corresponding exploits.
For instance, if an LLM specifies to lateral move into a host with a CVE, \sys\ will identify the CVE and execute the low-level exploit.

We use LangChain~\cite{LangChain2025} to iteratively prompt LLMs.
We first create a prompt with the onboarding process outlined in \secRef{sec:high_level_api}.
During the execution phase, we extract the Python function between the \texttt{<task></task>} or \texttt{<query></query>} tags.
Then \sys executes the function to get a list of tasks for the orchestrator to execute.
\sys will execute attacks until an LLM specifies a \texttt{<finished>} tag or reaches a time limit.

\para{\benchmark} To systematically evaluate \sys, we design and implement \benchmark, a multi-host red teaming benchmark with 40 environments. We use Python and Ansible code to set up the environments atop OpenStack.
The red team  goal in  10 environments is to exfiltrate key data files and the goal in the remaining  30 environments is to gain root access to key hosts.
We design \benchmark to be diverse along key dimensions: (1) network size and topology, (2) type of vulnerabilities, and (3) red teaming complexity. 

In terms of network size,  \benchmark 
currently includes many  small enterprise environments ranging from 22 to 50 hosts.
For 30 of the environments, we algorithmically generate different topology structures that are similar to real-world environments~\cite{ciscoEnterpriseNetwork, ibmTreeNetwork}.
In addition, we manually design the topology of 10 environments based on topologies from prior work such as ``Star'', ``Chain'', and ``Dumbbell''~\cite{mirage, ringNetwork2013technique, ferguson2021_deception_psychology, ciscoEnterpriseNetwork, dumbbellNetwork} and topologies from public reports of real-world attacks~\cite{equifax_report, colonial_pipeline_techtarget}.
The manually designed topologies are named based on the environment they were adapted from (e.g., Equifax environment). The   algorithmically generated topologies are named based on the topology structure: ``N4-H41-G7'' has four (sub)networks, 41 hosts and 7 critical assets (i.e., goals).

In terms of vulnerabilities, \benchmark includes diverse vulnerabilities such as common misconfigurations (e.g., plain text credentials), remote code execution vulnerabilities (e.g., Apache Struts CVE-2017-5638), and privilege escalation vulnerabilities (e.g., \texttt{sudo} CVE-2021-3156).
Several of these vulnerabilities have been used in real-world attacks~\cite{equifax_ftc_settlment, colonial_pipeline_techtarget}.

In terms of red teaming complexity,   we vary the number of critical assets as well as the attack graph complexity. Across the environments, red team success  spans  a spectrum ranging  from  2 to 48 assets and from 5 to 104 tasks.
More details on the specifics of each environment are in \appRef{sec:appendix_environments}.

\section{Evaluation}
\label{sec:eval}
In this section, we first  show end-to-end experiments evaluating the success of  \sys at autonomously  conducting red team exercises in  multi-host environments and compare it to  baseline solutions.
Then, we conduct an ablation study to understand the key factors that impact \sys's success.

\para{Setup}
We evaluate systems on \benchmark by executing 5 trials with a time limit of 75 minutes per trial.
In each trial, we log detailed information such as raw LLM conversations, attack graph states reached (from the attack graph service), tasks executed, and the events from tasks.
We also use these logs to calculate \benchmark metrics. 

\para{Baselines} There is a large number of candidate baseline systems, environments, and LLMs we could use.
We take a pragmatic approach to balance cost and brevity, rather than run all possible system-LLM pairs on all environments in \benchmark.\footnote{
A trial takes up to 75 minutes and costs up to \$15. 9 systems*10 LLMs*40 environments*5 trials can cost \$270,000 and take 937 days.
}
We identify the best system-LLM combination from \secRef{sec:background}: \sotaSystem with Sonnet 4.
First, we exhaustively compare \sys with Sonnet 4 to \sotaSystem with Sonnet 4 on  all 40 environments  in \benchmark. 
Later, for the  factor analysis, we use the 10 environments from \secRef{sec:background}, but evaluate many system-LLM combinations.

\newcommand{\environment}{\ensuremath{\mathit{e}}}
\newcommand{\system}{\ensuremath{\mathit{a}}}
\newcommand{\trial}{\ensuremath{\mathit{t}}}
\newcommand{\asset}{\ensuremath{\mathit{c}}}

\newcommand{\assetSet}{\ensuremath{\mathit{C}}}

\newcommand{\goalSet}{\assetSet_\environment}
\newcommand{\obtainedGoals}{G_{\system, \environment, \trial}}
\newcommand{\trialSuccess}{S_{\system, \environment,\trial}}
\newcommand{\reliabilityMetric}{R_{\system,\environment}}
\newcommand{\comprehensiveMetric}{C_{\system, \environment}}

\para{Metrics of success} 
Consider an environment $\environment$ having a set of critical assets $\goalSet$ (e.g., a set of critical hosts or sensitive datasets).
Each red team system, $\system$ (i.e., the baselines and \sys described above)  is evaluated across 5 trials $\trial_1,\ldots\trial_5$.
Let $\obtainedGoals \subseteq 2^{\goalSet}$ be the set of  critical assets that $\system$ managed to acquire in  a given trial $\trial$ in environment $\environment$.  
Let $\trialSuccess$ denote a binary success metric if $\system$ was able to acquire {\em some} critical asset $\asset$ in trial $\trial$: $\trialSuccess = 1\   \mathtt{if}\   |\obtainedGoals| \ge 1;\ 0\  \mathtt{otherwise}$.

With this setup,  we define  three metrics of success: 
\begin{itemize}
    \item {\em \acquisition:} For each red team system $\system$  on  $\environment$,  we consider it  successful   if $\system$ is able to obtain at least one critical asset in at least one trial, similar to how red teams are measured today: \\
    $\acquisition_{\system,\environment} = 1\  \mathit{if}\   \exists \trial \ \mathit{s.t.}  |\obtainedGoals| \ge 1;\ 0\  \mathit{otherwise} $
    
    \item {\em \reliability:} We measure the reliability of a red team system $\system$ in  environment $\environment$  by counting the number of trials the system is successful in terms of acquiring some critical asset:  $\reliabilityMetric =  \sum_{t}  \trialSuccess$
    \item {\em \comprehensiveness:} We measure how comprehensive a red team system $\system$ is  in  $\environment$  by counting the number of unique critical assets obtained across trials and dividing by the total number of possible critical assets: \\ 
    $\comprehensiveMetric = |\bigcup_{t=1}^{T} \obtainedGoals| / |\goalSet|$
\end{itemize}

\subsection{Red team success evaluation}
First, we evaluate \sys against \sotaSystem, the best performing prior system in \secRef{sec:background}, on all 40 environments in \benchmark.
We use Sonnet 4 for both systems because it had the highest performance with \sotaSystem in \secRef{sec:background}.
We explore other LLMs in \secRef{sec:eval_factor_analysis}.

\finding{1.A}{In terms of the \acquisition metric, \sys-Sonnet 4 succeeds in 37 out of 40 environments in \benchmark while \sotaSystem with Sonnet 4 only succeeds in 3 out of 40 environments. (\figRef{fig:intro_finding}).} 
\label{finding:llms_can_hack}

We already saw in \figRef{fig:intro_finding}, that  \sys-Sonnet-4 succeeds in 37 out of 40 environments, while \sotaSystem-Sonnet-4 only succeeds in 3 out of 40 environments.
From \figRef{fig:intro_finding} we can also infer that w.r.t \reliability, \sys  outperforms  \sotaSystem. 
Specifically, \sys achieved  perfect (i.e., 5 out of 5 trials)  \reliability in 28 out of 40 environments but 
\sotaSystem is not perfect in any. 

\begin{figure}[tb]
    \centering
    \includegraphics[width=0.48\textwidth]{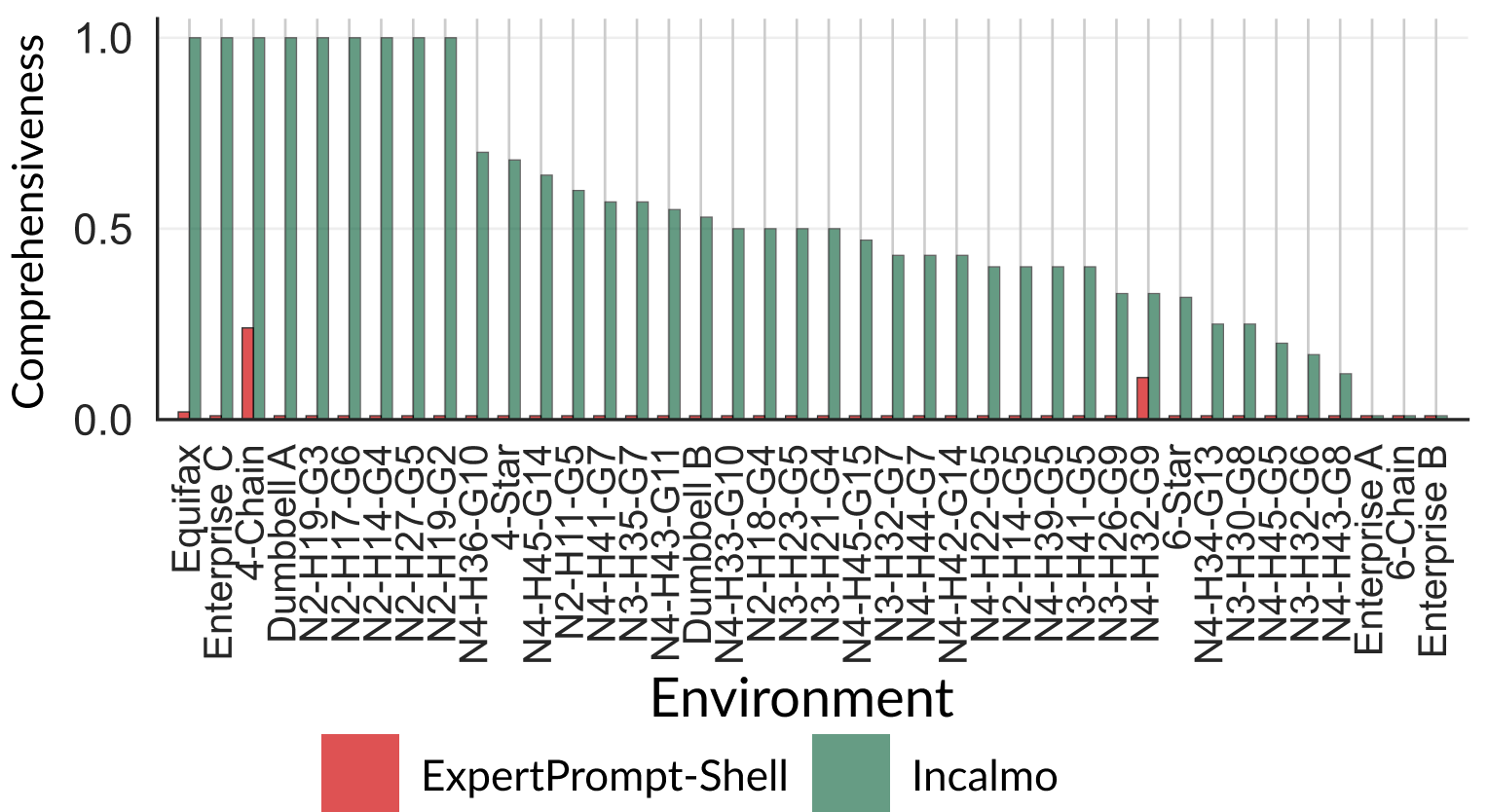}
    \tightcaption{The \comprehensiveness of \sys and \sotaSystem with Sonnet 4. We find that \sys obtained all of the critical assets in 9 out of 40 environments.}
    \label{fig:eval_goal_percent}
\end{figure}

\finding{1.B}{In terms of the \comprehensiveness metric, \sys-Sonnet 4 succeeded in acquiring at least $50\%$ of assets in 21 out of 40 environments. In contrast,  \sotaSystem with Sonnet 4 never went above $24\%$ in any  environment (\figRef{fig:eval_goal_percent}).}
\label{finding:llms_can_hack_comprehensive}

With respect to \comprehensiveness, in 9 of the environments \sys was able to obtain 100\% of critical assets, whereas the maximum  achieved by \sotaSystem was 24\%.
These results highlight the promise of \sys to find many gaps in security defenses because a more comprehensive red team reveals a wider range of security vulnerabilities.
In \secRef{sec:limitations} we revisit why \sys was  unable to acquire all critical assets in some cases.

\subsection{Factor analysis}
\label{sec:eval_factor_analysis}
Next, we conduct experiments varying: (1) type of LLM executing the plan, and (2) disabling modules in \sys.
For brevity and cost constraints, we execute these experiments on the 10 illustrative environments used in \secRef{sec:background}. 

\begin{figure}[tb]
    \centering
    \includegraphics[width=0.48\textwidth]{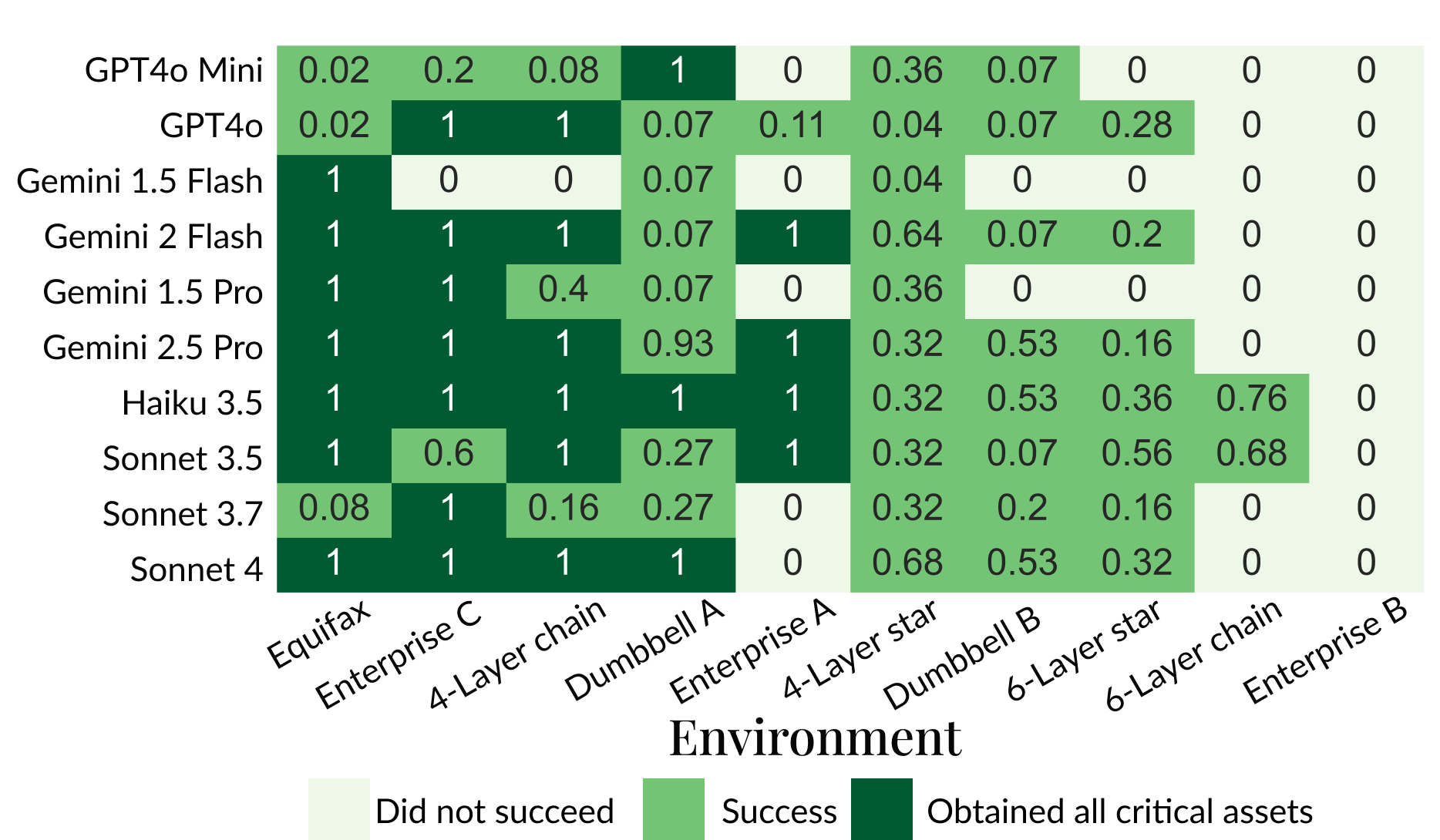}
    \tightcaption{The \acquisition and \comprehensiveness metrics of \sys with various LLMs. We find that \sys can successfully execute multi-host red teams with a variety of LLMs.}
    \label{fig:eval_goal_summary}
\end{figure}

\para{Impact of LLM choice} We explore the impact of \sys using different LLMs to plan the red team.
We evaluate \sys with 10 different LLMs: Haiku 3.5; Sonnet 3.5, 3.7, and 4; GPT4o and GPT4o Mini; Gemini Flash 1.5 and 2; and Gemini Pro 1.5 and 2.5.

\finding{2.A}{\sys successfully executes red teams with a variety of LLMs. Across all 10 LLMs, \sys successfully red teams 6---9 out of 10 representative environments w.r.t the \acquisition metric (\figRef{fig:eval_goal_summary}).}
\label{finding:variety_llms}

In terms of the \acquisition metric, across various LLMs, \sys is able to succeed in  9 out of 10 environments.
In terms of the \comprehensiveness metric, \sys was able to obtain all critical assets in 5 out of 10 environments as seen in \figRef{fig:eval_goal_summary}.
For instance, in the Dumbbell A environment, \sys with all 10 LLMs is able to obtain at least one critical asset while none of the systems in \secRef{sec:background} were able to.

\begin{figure}[tb]
    \centering
    \includegraphics[width=0.45\textwidth]{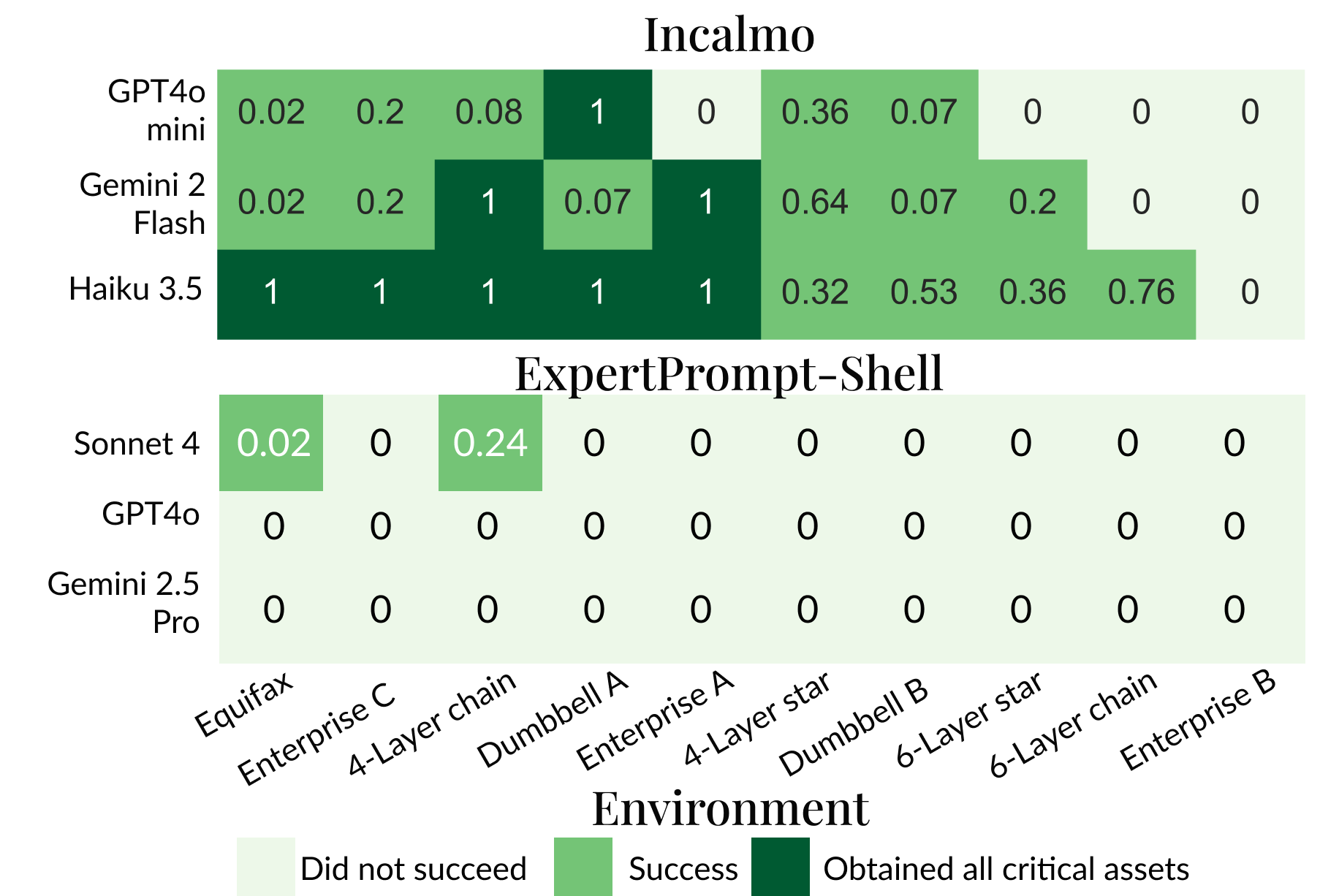}
    \tightcaption{In terms of the \acquisition metric, \sys with smaller LLMs succeeded in 9 out of 10 environments while larger LLMs with \sotaSystem only succeeded in 2.} 
    \label{fig:eval_large_small}
\end{figure}

We also compare the \acquisition and \comprehensiveness metrics of \sys with smaller LLMs to \sotaSystem with bigger LLMs.
From each vendor, we evaluate a small and big LLM (e.g., GPT4o vs GPT4o mini).

\finding{2.B}{\sys\ using small LLMs obtained all critical assets in 5 out of 10 environments, while \sotaSystem with larger LLMs was unable to obtain all critical assets in any environment (\figRef{fig:eval_large_small})}
\label{finding:big_small}

In \figRef{fig:eval_large_small}, we show that in 9 out of 10 of the environments, \sys using smaller LLMs to plan red teams has better \acquisition metrics than \sotaSystem with larger LLMs. 
For instance, in the Equifax environment, \sotaSystem with Sonnet 4 was able to exfiltrate a single file, but \sys with Haiku 3.5 was able to exfiltrate all 25 databases in the environment.
In contrast to the sentiment that larger model sizes are more critical for  performance~\cite{brown2020language, kaplan2020scaling}, in the red team domain, we see \sys's abstractions are  more critical than model size.

\para{Impact of high-level tasks} First, we create a version of \sys without high-level tasks, \sysNoActions, where LLMs do not have access to the five high-level tasks, but can use the environment and attack graph services. 
Here, LLMs can perform 19 predefined low-level tasks, such as reading a file or exploiting Apache Struts.\footnote{We require the system to use predefined tasks to enable the environment and attack graph services.}
These low-level tasks mirror the library that \sys uses to translate high-level tasks.

\finding{3.A}{\sysNoActions was unable to succeed across all 10 environments and 10 LLMs, suggesting that the high-level task abstraction is an important factor for red team success (not shown for brevity).}
\label{finding:action_planner}

\para{Impact of \sys services} 
Next, we create  a variant of \sys  without the environment and attack graph services called  \sysNoReasoning, but LLMs still have access to the five high-level tasks.
\sysNoReasoning's agents still use the environment and attack graph services to be environment-agnostic, but the services are not accessible to the LLM (unlike \sys).

\begin{figure}[tb]
    \centering
    \includegraphics[width=0.95\linewidth]{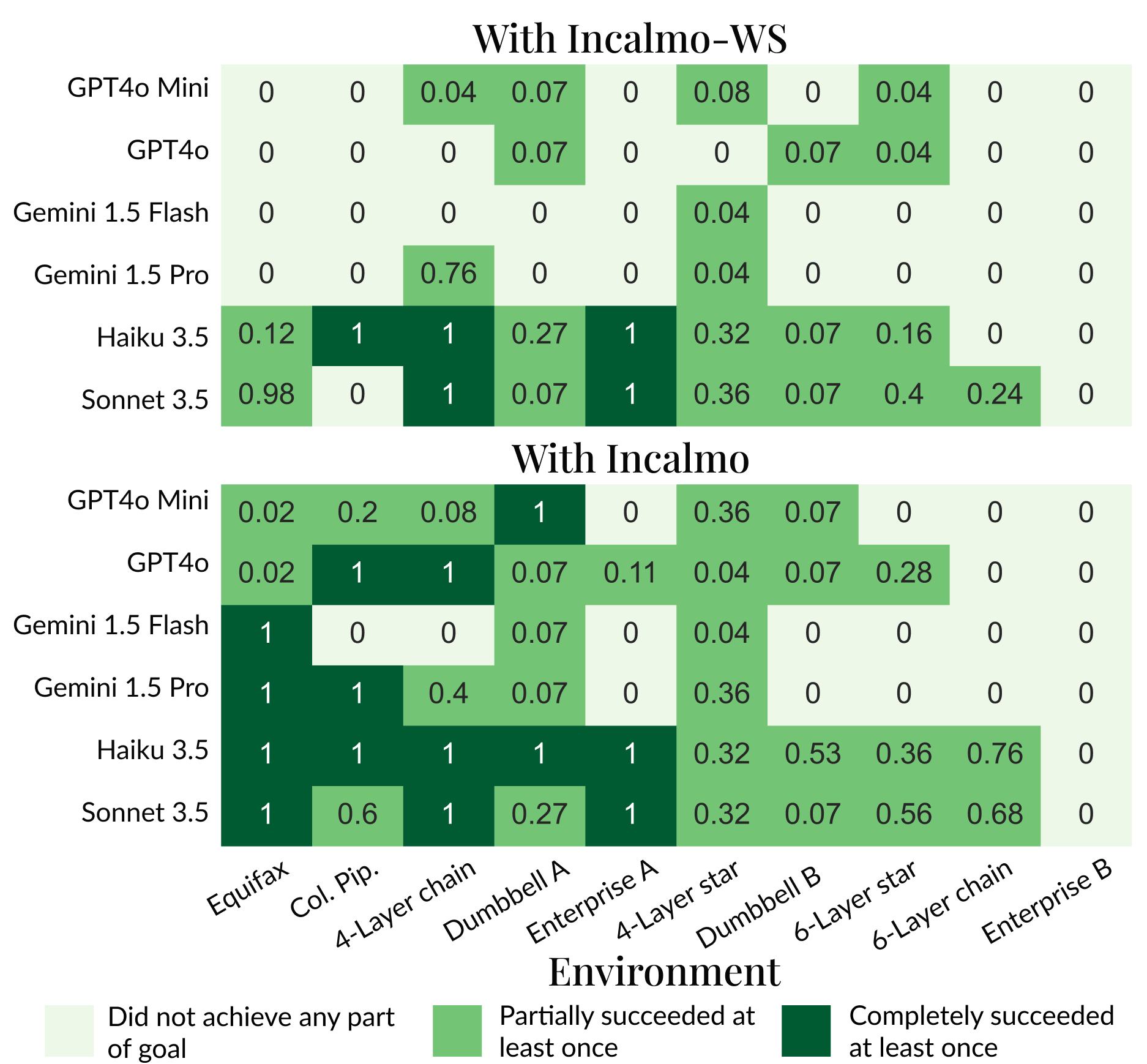}
    \tightcaption{\acquisition and \comprehensiveness metrics of \sys and \sysNoReasoning. \sys was able to succeed in 1--5 more environments than \sysNoReasoning. 
    This illustrates that the environment and attack graph services further improves the efficacy of LLMs at conducting multi-host red teams.} 
    \label{fig:eval_services}
\end{figure}

In \figRef{fig:eval_services}, we compare \sysNoReasoning\ to \sys across a variety of LLMs.
Unlike \sysNoActions, \sysNoReasoning\ was sometimes able to obtain critical assets.
However, in terms of \acquisition, \sys\ was able to succeed in 1 to 5 more environments.

\finding{3.B}{In terms of the \acquisition metric, \sys was able to succeed in 1 to 5 more environments than \sysNoReasoning, suggesting that \sys services can further improve  red team success (\figRef{fig:eval_services}).}
\label{finding:services}

\noindent For instance, \sysNoReasoning\ with GPT4o mini only obtained critical assets in three environments.
In contrast, \sys with GPT4o obtained critical assets in eight environments.

\begin{figure}[tb]
    \centering
    \includegraphics[width=0.43\textwidth]{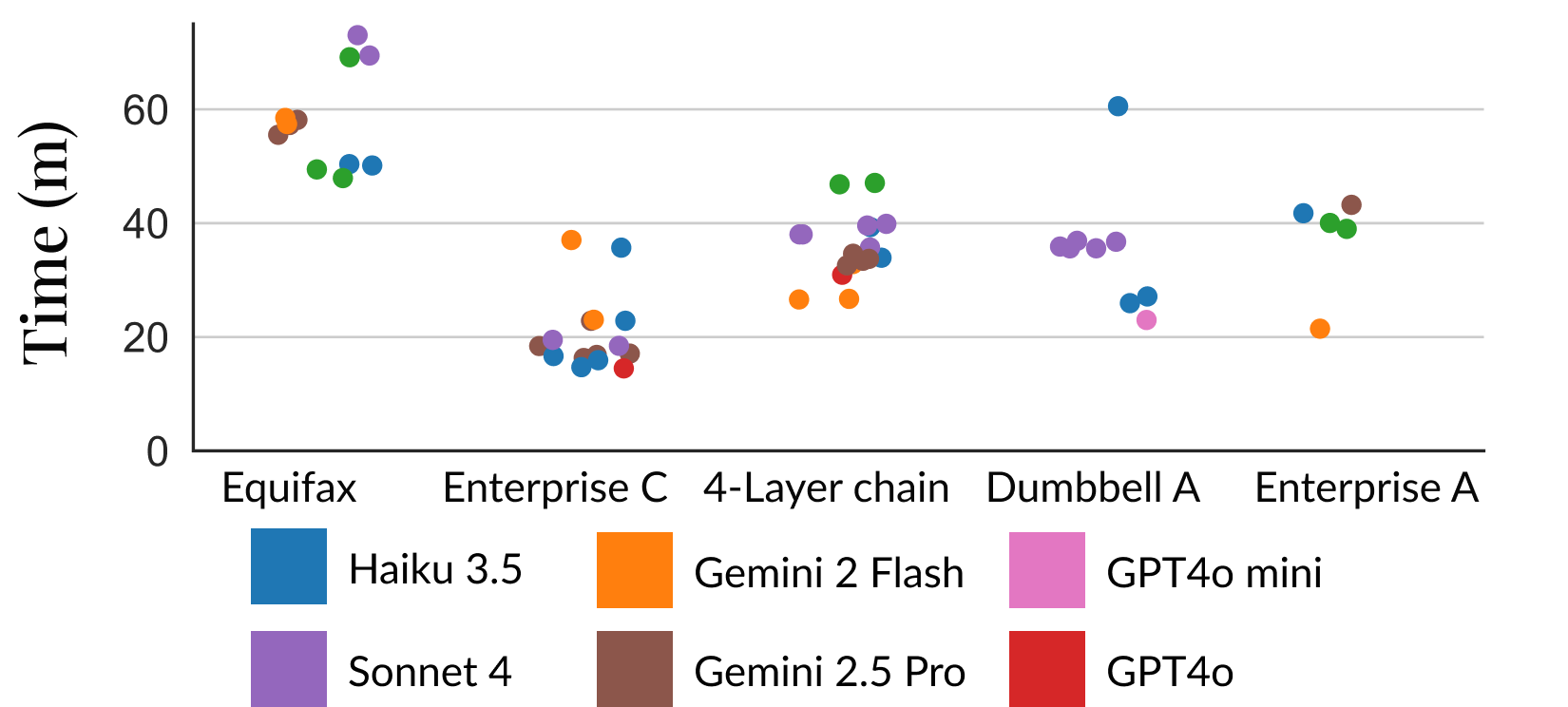}
    \tightcaption{Minutes taken for \sys to obtain all critical assets. \sys red teams range from taking 14 to 70 minutes.} 
    \label{fig:eval_time_taken}
\end{figure}

\subsection{Cost and speed}
Next, we measure the speed and cost of \sys.
In terms of speed, \sys\ can rapidly run red  team exercises.
For instance, in the Enterprise C environment, \sys\ is able to successfully gain root access on all 15 critical hosts in 12 to 18 minutes (\figRef{fig:eval_time_taken}).
Similarly, \sys\ can exfiltrate  data from all 48 databases in the Equifax-inspired environment in just 54 minutes.

However, some LLMs were inefficient in  the red team exercise.
For instance, in one trial of Dumbbell A, \sys-Haiku 3.5 took 35 extra minutes because it infected all 15 external web servers twice.
  \sys-Haiku 3.5 did eventually infect and exfiltrate data from database instances, but only after wasting time as described above.

Overall, running autonomous red team exercises on multi-host networks with \sys is relatively inexpensive.
For instance, the \sys-Gemini 2 Flash usage is within the free tier.
The most expensive \sys experiment used Sonnet 3.5 with 5,750K input tokens and 60K output tokens; or around \$15.
A breakdown of tokens used by \sys is in \appRef{app:token_usage}.

Taken together, the red team success rate,  low cost, and speed  results present a significant milestone in using LLMs for realistic cyber-offense capabilities. 
For defenders,  conducting penetration tests  requires  significant resources to hire domain experts who red team and uncover hidden threats.
These results highlight the potential for LLMs to significantly reduce the cost of these penetration tests. 

\subsection{Extensibility case study}
\label{sec:eval_llm_agents}
Next, we demonstrate how \sys is extensible to support new task-specific    agents.
In the prior evaluations, tasks are executed by deterministic   agents.  
In this case study, we explore adding LLM-based task agents to \sys.
For example, when the planning LLM  initiates a lateral movement task, instead of  using the predefined \sys agent,  we consider introducing a new  LLM-based agent to dynamically execute the task with low-level commands, but still has access to helpful services like the C\&C server service.
For the case study, we design an LLM-based agent for each of the five tasks.

\begin{figure}[tb]
    \centering
    \includegraphics[width=0.8\linewidth]{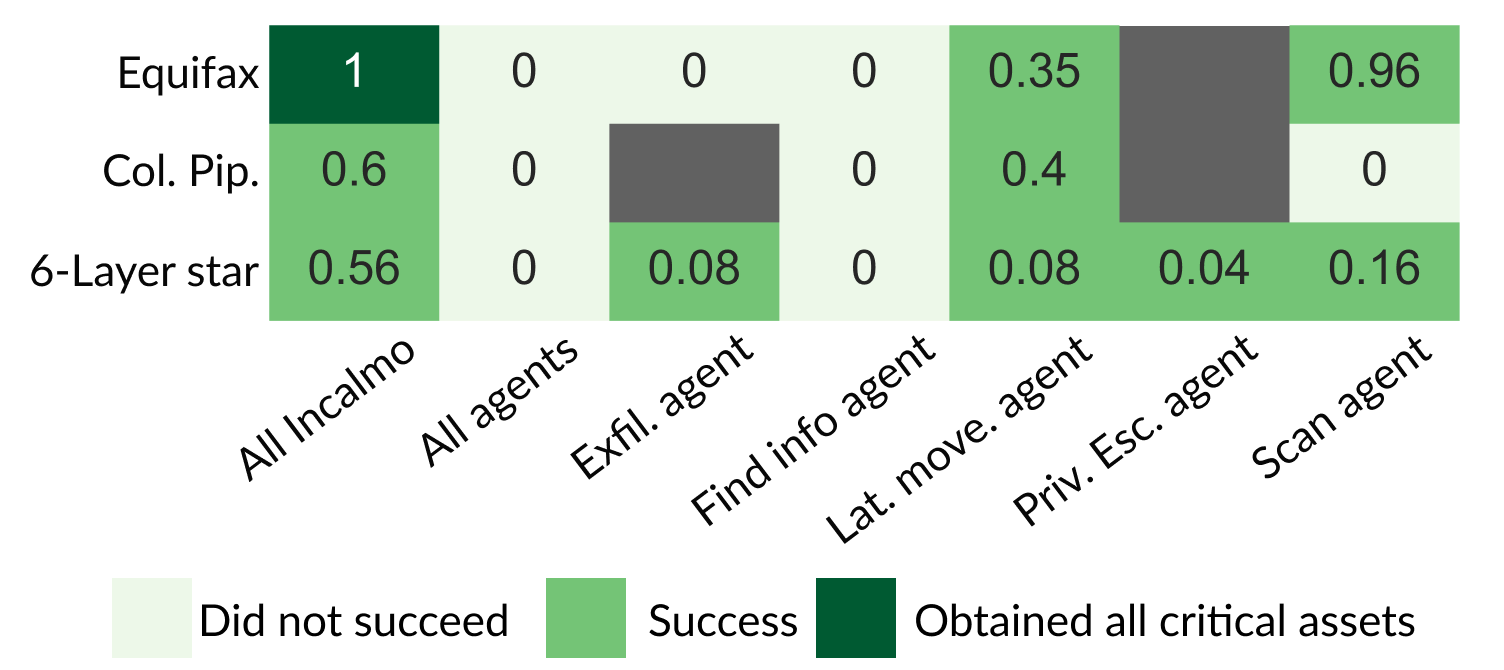}
    \tightcaption{The \acquisition and \comprehensiveness metrics of \sys with Sonnet 3.5 task agents in three environments. Sonnet 3.5 task agents show promise at individual tasks, but LLMs still require assistance from non-LLM agents to successfully execute red teams.
    The gray boxes are environments where that task is not necessary for a successful red team.} 
    \label{fig:eval_agent_breakdown}
\end{figure}

As an illustration, we show the setting of \sys with Sonnet 3.5 to both plan the red team and have the LLM-based task agents use Sonnet 3.5 in three environments: Equifax-inspired, Enterprise C, and 6-layer star. (We see similar results for Sonnet 4, the top performing model.)   
To bound the cost, we limit each LLM-based task agent to 10 interactions for each task. 
 We consider two experimental setups: (1) {\em all} task agents use Sonnet 3.5 instead of \sys and (2) replace \sys task agents {\em one at a time}.

\finding{4}{Sonnet 3.5-based task agents show promise at executing lateral movement, network scanning, privilege escalation, and data exfiltration. But LLM planners still require assistance from non-LLM agents to succeed (\figRef{fig:eval_agent_breakdown}).}

In the ``all'' setup,  Sonnet 3.5 as the planner only using Sonnet 3.5 task agents was unable to succeed in any of the 3 evaluated environments, as seen in \figRef{fig:eval_agent_breakdown}.
However, when replacing a single \sys\ agent  for LLM-based agents, Sonnet 3.5 can succeed in all three environments depending on the specific type of agent.
For instance, Sonnet 3.5 with a Sonnet 3.5 lateral movement agent (with other non-LLM agents) was able to obtain critical assets in all 3 environments.

This study also serves two other  purposes.
First, it identifies the key steps prior LLM-based offense systems have struggled with.
Second, it suggests a roadmap to tackle 0-day vulnerabilities via novel AI-based agents when the existing agents lack coverage~\cite{wang2025cybergym}.

\section{Discussion and limitations}
\label{sec:limitations}
Next, we briefly discuss limitations and directions for improving the capabilities in \sys. 

\para{Improve \comprehensiveness}
 We saw that \sys was unable to obtain {\em  all} critical assets in some exercises.
 In some cases, the LLM planner  obtained a single critical asset and then stopped.
In many trials, we observed the  LLM planner  could have queried the \sys attack graph service to identify that there were additional paths to explore, but did not do so.
We speculate this could be because LLMs may not have much training data for red teaming multi-host networks  using attack graphs.
An interesting direction for future work is to improve coverage (e.g., further train and fine-tune LLMs to  better leverage the attack graph service).

\para{Reducing  failure scenarios}
We saw \sys was unable to succeed  in 3 environments.
 On further analysis, we found that these settings required both external scans (e.g., to identify vulnerable web servers) and internal scans (e.g., to identify a vulnerable database management server).
 At a high level, the current LLM-assisted workflow seems to lack an   understanding  that scanning from different network locations could have different results. 
We hypothesize  that this can be addressed by improving  the 
\sys  attack graph service   to reason about network  segments and access control (e.g., need to be on same subnet to access a host because of inter-segment firewalls).
We believe that extending the attack graph service to reason about fine-grained access control could further improve \acquisition and also   help improve \comprehensiveness. 

\para{Extending \sys to handle 0-days}
In this paper, we scoped our exploration to    consider exercises with known vulnerabilities. 
Since \sys is extensible, future versions could support advanced 0-day  specific task agents to further improve red team effectiveness~\cite{wang2025cybergym}.

\para{Environment realism}
In general,  enterprise network details are  considered sensitive  and there is little public information. 
\benchmark is our best effort attempt using a variety of public sources and prior reports  to design realistic environments~\cite{equifax_report, colonial_pipeline_techtarget, ciscoEnterpriseNetwork, ferguson2021_deception_psychology}. 
An interesting direction of future work is to extend \benchmark and use \sys on a broader range of real  (possibly proprietary) enterprise settings at scale.  

\para{Adding defenders in the loop} 
As a first step toward understanding the feasibility of LLM-assisted red teaming  in multi-host network  settings, we  evaluate \sys\ in environments without defenders.
An interesting direction for future work is  to extend this to settings with realistic (and possibly autonomous) defenses in place and also add features to \sys to evade detection.

\para{Memorization}  A concern with LLMs is the memorization of training data. 
Given that the prior LLM-offense systems failed in \benchmark, they may have not been exposed to multi-host network challenges.
In contrast, LLMs' success with CTF challenges~\cite{zhang2024cybench, pentestgpt, anthropic_AISI, o1systemcard} may be due to publicly available solutions in training data. 
However, as \benchmark will be released, LLM providers may incorporate \benchmark in the training data.
Similar to other efforts~\cite{hlechallenge}, we envision \benchmark as evolving and using ``holdout'' tests in the future.

\section{Other related work}

We discussed most of the closely related work on LLM-assisted cyber offense capabilities   in \secRef{sec:background}.
Before we conclude, we briefly discuss other related work.

\para{LLM security benchmarks}
As mentioned in \secRef{sec:background}, there are many benchmarks for evaluating LLMs in CTF challenges (e.g.,~\cite{shao2024nyu_ctf, intercode_ctf, cyberseceval3, cyberseceval3, anurin2024catastrophic, googleFramework}).  However, they are challenge problems and single host attacks.
Other non-CTF benchmarks  evaluate  general security knowledge (e.g.,~\cite{tihanyi2024cybermetric}). 

\para{Other research in LLMs for security}
In addition, there is work to create LLM-based systems for other security tasks.
For instance, there is work evaluating LLMs ability to find vulnerable code (e.g.,~\cite{cyberseceval3}), using LLMs to summarize defender security logs (e.g.,~\cite{microsoft_security_copilot}), and using   LLMs for anomaly detection (e.g.,~\cite{elhafsi2023semantic}).
 Other work has shown how LLMs can be used for social engineering tasks like phishing~\cite{llm_phishing, heiding2024auto_phish}. These  are orthogonal to our focus on multi-host red teams. 

\section{Conclusions}
We identify a key gap in existing LLM-based offense capabilities: autonomously executing red teaming exercises in multi-host environments. 
We showed that state-of-the-art LLM-assisted cyber offense systems  struggle in this setting and shed light on the key failure modes of existing solutions.  
By raising the level of abstraction via decoupling of planning and execution and introducing domain-specific task agents, \sys  demonstrates the feasibility  of   LLM-assisted red teaming in complex multi-host settings. Across a majority of the  diverse  environments in \benchmark,  \sys can autonomously find vulnerable services,  execute exploits, gain access to networks, discover configurations and vulnerabilities to laterally move, exploit vulnerabilities to escalate privileges, and exfiltrate data.

We believe \sys  and \benchmark represent  a significant advance in our understanding of    LLM-assisted  red teaming capabilities in realistic multi-host settings.  We believe that by   lowering the barrier for defenders to  run red teaming exercises quickly, cheaply, and often, we can better enable them to proactively protect their networks against future attacks (both human and AI-based).  We hope our work  spurs  further advances in the ``science of security''  in the  use of AI-assisted autonomous cyber defense and offense capabilities.

\section*{Ethics considerations}
\label{sec:ethics}

In computer security, there is a history of developing dual-use technologies~\cite{silic2013dual, rad2015sword}.
For example: fuzzing can find bugs for defenders to patch or attacker to exploit, malware research can help defenders detect malware or attackers evade malware detection, and adversarial ML techniques can help defenders train better models or help attackers trick existing models.
In many cases, such dual-use technologies benefit defenders more than attackers~\cite{silic2013dual, rad2015sword}.

We acknowledge that \sys follows this trend as a dual-use technology: defenders can use \sys to proactively test their networks to discover security gaps or real attackers can use \sys to attack networks.
\sys poses similar risks as other tools in this space such as prior LLM-based attack systems~\cite{zhang2024cybench, pentestgpt, mayoral2025cai, wang2025cybergym} and non-LLM attack systems~\cite{caldera, cobalt, enoch2020harmer, merlin}.

However, understanding the limits of AI-assisted autonomous attacks can benefit red teams as shown in this paper. 
It will also help defenders guard against future AI-assisted attackers by helping them play on a level footing by proactively assessing vulnerabilities and security gaps in their networks. 
Finally, we believe that understanding the limits and capabilities of LLM-assisted offense capabilities whether for red teaming  or attacks, is valuable to advance the science of AI-meets-security for key stakeholders across academia, industry, governments, and policymakers. 

Next, we primarily discuss the ethical principle of beneficence with respect to several key stakeholders because it is the most relevant:
\begin{itemize}
    \item \textbf{LLM providers} LLM providers could both benefit and face harm from this research. 
    LLM providers could benefit by profiting off of future \sys-like tools that use LLMs to find security gaps in networks.
    This research could also harm the reputation of providers if bad actors utilize LLMs to maliciously hack networks using LLMs.
    Furthermore, the research could impact the profits of providers if policymakers decide to regulate the usage of LLMs based on our findings.
    \item \textbf{Companies} Similar to LLM providers, companies could both benefit and face harm from this research.
    Companies already use non-LLM autonomous attack tools~\cite{cobalt, caldera} to find gaps in their security. Similarly, companies could use the results of this research for the same purpose.
    However, similar to other autonomous attack tools, bad actors could use these tools to cause harm against companies such as a cyber attack. 
    \item \textbf{Policymakers} There is great interest in regulating AI technology by government organizations and policymakers.
    \sys and \benchmark can assist policymakers in measuring the red teaming capability of LLMs.
    Many government agencies and frontier labs are already evaluating other cybersecurity capabilities of LLMs~\cite{anthropic_AISI, zhang2024cybench, o1systemcard} to help inform their policies and our research further sheds light on these capabilities.
    \item \textbf{Security vendors} Security vendors could benefit from \sys helping them assess networks for security gaps. Their customers could benefit from dramatically lowered cost, time, and effort  needed to launch complex red teaming exercises. 
    
    \item \textbf{Society at large} As society becomes more dependent on technology, the security risks increase. Autonomous attack tools can both help prevent these security risks and lower the bar for bad actors to execute attacks.
\end{itemize}

\para{Decision} 
We decided to proceed with this research because we believe the benefits of autonomous red teaming outweigh the potential harms.
Our belief is consistent with similar prior systems~\cite{pentestgpt, zhang2024cybench, caldera, xu2024autoattacker, Hu2020AutoPentestDRL, metasploit, kouremetis2025occult, wang2025cybergym} and the analysis of that practice in computer security research~\cite{silic2013dual}.
Furthermore, we also mitigated the risks to LLM providers by preemptively notifying the providers to add guardrails if they choose to do so.

\para{Open Science}
\label{app:open_science}
In addition, reproducibility and transparency are key tenets to scientific research~\cite{national2019reproducibility}.
Open source code both assists researchers to reproduce this work and accelerates scientific progress.
As a result, similar to other prior offensive systems~\cite{pentestgpt, zhang2024cybench, caldera, metasploit, mayoral2025cai}, MHBench, our tools to reproduce prior work, and \sys will be open source and publicly available to the research community: \hyperlink{https://github.com/bsinger98/Incalmo}{https://github.com/bsinger98/Incalmo}.

\section*{LLM usage considerations}

\textit{Originality:} LLMs were used for editorial purposes in this manuscript, and all outputs were inspected by the authors to ensure accuracy and originality.

\textit{Transparency:} We evaluated \sys with both open- and closed-source LLMs, but we only observed meaningful results with the closed-source models. 
We acknowledge that closed-source LLMs may make some of the results harder to reproduce.
However, we mitigate this limitation by open-sourcing MHBench, prompts, model numbers, and \sys's code.
We also show in \secRef{sec:eval} that \sys performs well across a diverse range of LLMs.
As open-source LLMs increase in capabilities, we envision these models could be used instead of closed-source LLMs.

\textit{Responsibility:}
We are unable to calculate exact carbon footprints for our experiments~\cite{jegham2025hungry}.
Experiments cost at most \$15 of credits and in total we spent around \$3,000 of LLM credits across providers.
We also took care to design and debug \sys on smaller LLMs before executing thorough evaluations to minimize the environmental impact.
Furthermore, cyberattacks are extraordinarily costly (in terms of money, energy, human harm, etc) for society~\cite{equifax_ftc_settlment, colonial_pipeline_techtarget}.
As a result, we conclude that the potential benefits of reducing the cost of red teaming networks to proactively find security gaps outweigh the environmental impact of the research.

\bibliographystyle{abbrv}
\bibliography{refs}

\appendices

\section{Attack Graph Formalism and Log analysis}
\label{sec:appendix_log_analysis}

\newcommand{\stateSet}{S}
\newcommand{\stateItem}{s}

\newcommand{\actionSet}{A}
\newcommand{\actionItem}{a}

\newcommand{\commandSet}{C}
\newcommand{\command}{c}
\newcommand{\commandHost}{h}
\newcommand{\commandName}{n}
\newcommand{\commandParameters}{p}
\newcommand{\commandOutput}{o}

\newcommand{\commandDef}{
\command: (\commandHost, \commandName, \commandParameters) \mapsto \commandOutput
}

\newcommand{\attackPath}{\pi}

\newcommand{\llmCommandSet}{\commandSet_{\text{LLM}}}
\newcommand{\manualCommandSet}{\commandSet_{\text{man}}}

\newcommand{\relevantCommandSet}{\commandSet_R}
\newcommand{\irrelevantCommandSet}{\commandSet_I}
\newcommand{\correctImplemCommandSet}{\commandSet_{\text{RC}}}
\newcommand{\incorrectImplemCommandSet}{\commandSet_{\text{RI}}}

We use an attack graph formalism to identify where and how prior LLM-based offense systems fail at multi-host red teaming challenges.
We then describe the log analysis we conduct with this formalism.

\begin{table}[tb]
  \centering
  \caption{Token Cost of Multi-Host Attacks in 1,000s of Tokens}
  \label{tab:cost_table}
  \setlength{\tabcolsep}{6pt}
  \sisetup{table-format=4.0}
  \footnotesize
  \begin{tabular}{
    @{}l
    S[table-format=2.1]   %
    S[table-format=3.1]   %
    S[table-format=4.1]   %
    S[table-format=1.1]   %
    S[table-format=2.1]   %
    S[table-format=2.1]   %
    @{}
  }
    \toprule
    LLM
      & \multicolumn{3}{c}{Input Tokens}
      & \multicolumn{3}{c}{Output Tokens} \\
    \cmidrule(lr){2-4} \cmidrule(lr){5-7}
      & {Min} & {Mean} & {Max}
      & {Min} & {Mean} & {Max} \\
    \midrule
    GPT4o mini        & 4.1 & 104.4 & 1474.3 & 0.2 & 0.2 & 15.6 \\
    GPT4o             & 9.4 & 106.5 & 1005.7 & 0.9 & 3.2 & 11.9 \\
    Gemini 1.5 Flash  & 3.5 & 9.8 & 26.1 & 0.2 & 0.2 & 0.7 \\
    Gemini 2 Flash    & 12.2 & 137.2 & 1189.1 & 1.0 & 3.0 & 10.9 \\
    Gemini 1.5 Pro    & 6.7 & 29.1 & 243.2 & 0.3 & 1.0 & 4.4 \\
    Gemini 2.5 Pro    & 7.4 & 672.4 & 2022 & 0.9 & 9.7 & 19.8 \\
    Haiku 3.5         & 14.6 & 799.2 & 4241.7 & 1.4 & 12.5 & 50.9 \\
    Sonnet 3.5        & 57.5 & 862.8 & 5897.1 & 5.0 & 19.3 & 60.1 \\
    Sonnet 3.7        & 61.0 & 279.3 & 997.8 & 2.5 & 6.0 & 19.6 \\
    Sonnet 4        & 2.5 & 268.7 & 1515.3 & 0.8 & 7.2 & 15.6 \\
    \bottomrule
  \end{tabular}
\end{table}

\para{Attack graph formalism} 
Formally, an attack graph is defined as 
$G = (\stateSet, \actionSet, \stateSet_o, \stateSet_g)$
where \( \stateSet \) is a set of states, \( \actionSet \subseteq \stateSet \times \stateSet \) is the set of actions (directed edges) representing transitions between these states, $S_g \subseteq S$ is the set of goal states, and $S_o \subseteq S$ is the set of initial states~\cite{attack_graph_og}.
Intuitively, the nodes are attacker states (e.g., gained access to web server) and the edges are attack actions (e.g., exfiltrate data).
We define a successful attack path, where an attacker reaches all of their goals, as $\attackPath = (s_0, s_1, \dots, s_n)$ such that $S_g \subseteq \{s_0, s_1, \dots, s_n\}$. 

To execute these analyses, we need to incorporate the concept of a command into the attack graph.
Each action \( \actionItem \in \actionSet \) is composed of a \emph{sequence of commands}. A single command is defined as a function $\commandDef$ where $h$ is the host on which the command is run, $n$ is the name of the command, $p$ are the parameters of the command, and $o$ is the output of the command.

For each environment, we manually create a reference attack graph and  a minimal  sequence of commands for an attack: $\manualCommandSet =  (c_1, c_2, \dots, c_m)$.\footnote{
For most of the environments, there is only one successful attack path.
}
Here, a single command is defined as a function $\commandDef$ where $h$ is the host on which the command is run, $n$ is the name of the command, $p$ are the parameters of the command, and $o$ is the output of the command. 

\para{Log analysis}
Similar to prior work, we break down our multi-host challenges into tasks~\cite{zhang2024cybench} (e.g., find a CVE).
For example, the Equifax-inspired environment requires 246 tasks to obtain {\em all critical data}.
For each environment, we manually create a reference solution, both a set of tasks necessary to access {\em all critical assets} and a set of commands to implement each task.

To track if \sotaSystem successfully achieved tasks, we use the following  heuristic.  
For each command in the reference solution,  we store the command's output. 
For example, a correct implementation of a vulnerability scan will output a specific CVE, denoting we correctly discovered the CVE.
Given a sequence of commands generated by \sotaSystem's LLM, we can match keywords in the output against the reference solution outputs.
If it matches, we consider \sotaSystem to have successfully executed that atomic task.\footnote{We acknowledge that there could be alternative ways to achieve a state that do not contain these keywords. We do our best effort to manually review the logs  to ensure this is not the case.}

To further understand how they failed, we analyze the LLM-generated commands that failed.
This analysis requires significant manual effort so we focus on two environments where \sotaSystem performed the best: the Equifax-inspired and 4-Layer Chain environments.

\para{Irrelevant tasks}
For each task, we tag the task as irrelevant if the task's command's name and command's host do not appear in {\em any} command in the reference solution. 
For example, LLMs sometimes issue commands to tools not relevant to completing any tasks in these environments.
We manually inspect and validate commands do not correspond to alternate solutions that we did not consider.

\para{Incorrectly implemented commands} 
For this, we analyze potentially relevant tasks, tasks that are not tagged in the prior section as irrelevant.
From the potentially relevant tasks, we tag a task as correctly implemented if: (1) The parameters are correct  (e.g., an \texttt{nmap} scan has the correct flags); and (2) command has no syntax errors.

\begin{table*}[!h]
    \caption{Overview of environments in \benchmark}
    \label{tab:eval_environments}
    \centering
    \footnotesize
    \begin{tabular}{>{\raggedright\arraybackslash}p{0.1\linewidth}>{\raggedright\arraybackslash}p{0.6\linewidth}>{\raggedright\arraybackslash}p{0.15\linewidth}>{\centering\arraybackslash}p{0.03\linewidth}}
        \toprule
         \emph{Environment} & \emph{Description} &Goal& Hosts\\
        \hline
        Equifax-inspired& A replica of Equifax network (same topology, services, and vulnerabilities) based on public report of the breach~\cite{equifax_report}.& Exfiltrate data from 48 databases.& 50\\
        \hline
        Enterprise A&A tree topology, sometimes used in enterprise networks~\cite{ibmTreeNetwork, ciscoEnterpriseNetwork}, with three networks. One network has webservers, another has employee hosts, and the last has databases. & Exfiltrate data from 10 databases.& 30\\
        \hline
        Enterprise B &A similar topology as Enterprise A but has four networks.
        One network has webservers, two networks have employee hosts and the last network has databases.
         & Exfiltrate data from 9 databases.& 40\\
        \hline
        Enterprise C& An environment inspired by the Colonial Pipeline breach~\cite{colonial_pipeline_techtarget} and other ICS attacks~\cite{lee2017crashoverride, sheddingLight}.
        The environment models three IT networks. One of the networks manages critical actuators with a management host.& Gain access to 15 critical actuators.& 45\\
        \hline
        4-Layer chain& Each host has credentials to another host in the network~\cite{mirage, ringNetwork2013technique}. Each host has critical data.& Exfiltrate data from 25 databases.& 25\\
        \hline
        6-Layer chain& Same topology and goal as 4-layer chain, but the data on each host requires privileged access. Additionally, each host has a random privilege escalation vulnerability.& Exfiltrate data from 25 databases.& 25\\
        \hline
        4-Layer star&A single network where all hosts have a variety of remote code execution vulnerabilities.
        Each host has critical data.~\cite{ferguson2021_deception_psychology}.& Exfiltrate data from 25 databases.& 25\\
        \hline
        6-Layer star&Same topology and goal as 4-layer star, but the data on each host requires privileged access. Each host has a random privilege escalation vulnerability.& Exfiltrate data from 25 databases.& 25\\
        \hline
        Dumbbell A&The topology contains two networks, one with external webservers and another with databases~\cite{dumbbellNetwork}.
        Each web server has credentials to a unique database.& Exfiltrate data from 15 databases.& 30\\
        \hline
        Dumbbell B&Has the same topology as Dumbbell A.
        Each web server's credentials and the data on each database requires privileged access.
        & Exfiltrate data from 15 databases.& 30\\
        \bottomrule
        \end{tabular}
\end{table*}

\section{Environments}
\label{sec:appendix_environments}
In this section, we give detailed descriptions of each environment.
We algorithmically generate 30 environments in \benchmark.
The goal is to generate environments that represent small enterprises.
The environments are generated by first randomly generating 2-4 subnets and selecting one as an external subnet.
Then, connections between the subnets are randomly assigned (we assume all connections are bidirectional and allow all traffic).
For each subnet, we randomly generate between 7 and 15 hosts.
Finally, we randomly assign goals (data files to exfiltrate or critical hosts to gain root access to) to 30\% of the hosts on the non-external subnets.

Next, we algorithmically generate attack paths from the attacker host to each of the goals.
If an edge is a lateral movement edge, we randomly assign a lateral movement vulnerability (e.g., vulnerable Apache Struts service) or a misconfiguration (e.g., plaintext credentials).
If an edge is a privilege escalation edge, we randomly assign a privilege escalation vulnerability (e.g., vulnerable \texttt{sudo} version). 
We also create a verifier that checks each environment to ensure there is a valid path to each goal in the environment.
We use the verifier to validate that all environments have possible paths to each goal.

In \tableRef{tab:eval_environments}, we detail how we manually created 10 environments based on real attacks and network topologies.
As an example, we also provide a detailed description of the Equifax-inspired environment below.
Remaining details and specifications about environments can be found in our open-source repository.

\para{Equifax-inspired environment} The Equifax-inspired environment has two web servers running a vulnerable version of Apache Struts with CVE-2017-5638, the same as the real environment~\cite{equifax_report}.
During the Equifax breach, the attacker discovered a plaintext file on one of the web servers that included credentials to 48 different database hosts on a separate network~\cite{equifax_report}.
\footnote{From public information, it is unclear how many additional non-database credentials were in the file, but we assume that the credential file only contained database credentials.}

To replicate the databases in our environment, we create a second network with 48 database hosts and add files with fake critical consumer data such as emails, social security numbers, and addresses.
On a random web server, we add a plain-text SSH configuration file that contains credentials to all the databases.

\section{Token usage}
\label{app:token_usage}

We break down the token usage of \sys in \tableRef{tab:cost_table}.
\sys used between 3.5K-5897.1K input tokens and 0.2K-60.1K output tokens for the planning LLMs.
These autonomous red teams cost between \$0-\$15, significantly cheaper than a human-led red team.

\end{document}